# Two-dimensional probabilistic inversion of plane-wave electromagnetic data: Methodology, model constraints and joint inversion with electrical resistivity data

Marina Rosas Carbajal[1], Niklas Linde[1], Thomas Kalscheuer[2] and Jasper A. Vrugt[3,4]

[1]Applied and Environmental Geophysics Group, Faculty of Geosciences and Environment, University of Lausanne, 1015 Lausanne, Switzerland.

[2]Institute of Geophysics, ETH Zurich, Zurich, Switzerland.

[3]Department of Civil and Environmental Engineering, University of California Irvine, 4130 Engineering Gateway, Irvine, CA 92697-2175, USA.

[4]Institute for Biodiversity and Ecosystems Dynamics, University of Amsterdam, Amsterdam, The Netherlands.

**SUMMARY**

Probabilistic inversion methods based on Markov chain Monte Carlo (MCMC) simulation are well suited to quantify parameter and model uncertainty of nonlinear inverse problems. Yet, application of such methods to CPU-intensive forward models can be a daunting task, particularly if the parameter space is high dimensional. Here, we present a two-dimensional (2D) pixel-based MCMC inversion of plane-wave electromagnetic (EM) data. Using synthetic data, we investigate how model parameter uncertainty depends on model structure constraints using different norms of the likelihood function and the model constraints, and study the added benefits of joint inversion of EM and electrical resistivity tomography (ERT) data. Our results demonstrate that model structure constraints are a necessity to stabilize the MCMC inversion results of a highly-discretized model. These constraints decrease model parameter uncertainty and facilitate model interpretation. A drawback is that these constraints may lead to posterior distributions that do not fully include the true underlying model, because some of its features exhibit a low sensitivity to the EM data, and hence are difficult to resolve. This problem can be partly mitigated if the plane-wave EM data is augmented with ERT observations. The hierarchical Bayesian inverse formulation introduced and used herein is able to successfully recover the probabilistic properties of the measurement data errors and a model regularization weight. Application of the proposed inversion methodology to field data from an aquifer demonstrates that the posterior mean model realization is very similar to that derived from a deterministic inversion with similar model constraints.

## 1. Introduction

Geophysical measurement methods make it possible to non-invasively sense the physical properties of the subsurface at different spatial and temporal resolutions. Inversion methods are required to interpret these indirect observations and derive a physical description of the subsurface, yet multiple descriptions can be found (also referred to as models) that fit the observed geophysical data equally well. This is in large part due to measurement errors, incomplete data coverage, the underlying physics and/or over-parameterization of the subsurface models. Whereas the probabilistic properties of observation errors are relatively easy to describe, model structural errors are difficult to formulate in probabilistic terms. Arbitrary and subjective regularizations and parameterizations may significantly decrease model parameter uncertainty but they may also introduce a “bias”, meaning that some features of the true model may not be resolved.

Bayesian inference can help to explicitly treat input data, parameter, and model uncertainty, but successful implementation requires efficient sampling methods that explore the posterior target distribution. In this probabilistic approach, the inverse problem is stated as an inference problem where the solution is given by the posterior probability density function (pdf) of the model parameters. This distribution quantifies joint and marginal parameter uncertainty. Unfortunately, in most practical applications, this posterior distribution cannot be derived analytically, and methods are required that use trial-and-error sampling to approximate the target distribution. Markov chain Monte Carlo (MCMC) simulation methods are well suited for this task, but suffer from poor efficiency, particularly when confronted with significant model nonlinearity, nonuniqueness and high-dimensional parameter spaces (Mosegaard & Tarantola 1995).

The basic building block of MCMC sampling is Monte Carlo (MC) simulation. This approach randomly samples the prior parameter space, and evaluates the distance of the response of each candidate model to the respective data. If the parameter space is low dimensional, MC simulation can provide a reasonable approximation of the posterior distribution pending that the ensemble of samples is sufficiently large. Yet, for higher dimensional spaces, exhaustive random sampling is inefficient, and more intelligent search methods such as MCMC simulation are required to speed up the exploration of the target distribution. Monte Carlo methods have been applied to magnetotelluric (MT) data and other types of frequency-domain electromagnetic (FDEM) data in a number of studies for 1D modeling problems (Tarits *et al.* 1994; Grandis *et al.* 1999; Grandis *et al.* 2002; Khan *et al.* 2006; Hou *et al.* 2006; Chen *et al.* 2007; Guo *et al.* 2011; Minsley 2011; Buland & Kolbjornsen 2012). We briefly summarize a few of these studies.

Tarits *et al.* (1994) used Monte Carlo sampling to estimate the posterior distribution of the thicknesses and electrical resistivity of different subsurface layers assuming that the number of layers is known *a-priori*. Grandis *et al.* (1999) extended this 1D approach by employing MCMC simulation with sampling from a prior distribution that favors smooth variations in the 1D electrical resistivity model. Hou *et al.* (2006) used a quasi-Monte Carlo method (Ueberhuber, 1997, p. 125) for 1D models of reservoir-fluid saturation and porosity to jointly invert controlled source electromagnetic (CSEM) and seismic data. The same types of data were jointly inverted by Chen *et al.* (2007) using MCMC simulation to derive 1D models of gas saturation.

In a more recent contribution, Guo *et al.* (2011) compared deterministic and Bayesian MT data inversion using 1D synthetic and field data. Data errors and regularization weight were treated as hyper-parameters and determined by MCMC simulation (c.f. Malinverno & Briggs 2004). Results showed that the MT data contained sufficient information to accurately determine these latent variables. Minsley (2011) presented a 1D trans-dimensional MCMC inversion (Malinverno 2000) algorithm for FDEM data, in which the number of layers was assumed unknown. Their approach favors model parsimony between models that equally fit the data. This favoring of simple models is naturally accounted for in the so-called "Ockham factor", which measures how much of the prior information is contained in the posterior probability density function (pdf). With increasing number of parameters, the probability mass of the prior in the vicinity of the posterior will typically decrease (and so will the Ockham factor), while the data fit will typically improve (Malinverno 2002). Ray & Key (2012) used the same type of method to determine 1D anisotropic resistivity profiles from marine CSEM data. Most recently, Buland & Kolbjornsen (2012) jointly inverted synthetic CSEM and MT data and presented a real-world application for CSEM data. Khan *et al.* (2006) used EM data within a MCMC framework to constrain the composition and thermal state of the mantle beneath Europe.

The published contributions summarized thus far have demonstrated the ability of MCMC methods to (1) successfully converge to the global optimum of the parameter space, (2) treat nonlinear relationships between model and data, and (3) adequately characterize parameter and model uncertainty. Yet, all these studies used relatively simple 1D models to minimize the computational costs of the forward solution, and considered relatively low-dimensional parameter spaces to facilitate convergence of the MCMC sampler to the appropriate limiting distribution.

Grandis *et al.* (2002) presented the first published multi-dimensional MCMC inversion of MT data using a thin-sheet modeling code that is CPU-efficient, but only accurate for relatively thin anomalous bodies. Inversions were presented for a horizontal 2D anomaly embedded in a known horizontally layered 1D model. Chen *et al.* (2012) presented a MCMC algorithm to invert 2D MT data. They fixed the number of layers in the model, yet allowed the depths to vary at given offsets. A 2D resistivity structure was estimated at a geothermal site using 436 model parameters. This particular algorithm enables the inversion of 2D data, but imposes strict constraints on the model parameterization in that only layered models with sharp boundaries are allowed.

Other global search methods of stochastic nature, such as simulated annealing (Kirkpatrick *et al.* 1983) and genetic algorithms (Holland 1975), have been used to produce 1D and 2D electrical resistivity models from MT data (Dosso & Oldenburg 1991; Everett & Schultz 1993; Pérez-Flores & Schultz 2002). These methods fully account for the nonlinear relation between model and data, but are only concerned with finding the optimal model, without recourse to estimating the underlying posterior parameter distribution. Post processing of the sampled trajectories can provide some insights into the remaining parameter uncertainty, but this type of analysis approach lacks statistical rigor.

More complex and highly parameterized 2D or 3D resistivity models are generally obtained through deterministic inversion (e.g., deGroot-Hedlin & Constable 1990; Rodi & Mackie 2001; Siripunvaraporn & Egbert 2000; Siripunvaraporn *et al.* 2005). These algorithms are much more efficient but provide only a single "best" solution to the inverse problem (e.g., Menke 1989). Approximate uncertainty estimates can be obtained through linearization in the vicinity of the final solution (Alumbaugh & Newman 2000). As an alternative to such approaches, Oldenburg & Li (1999) derived a set of different deterministic models using the same data set by running repeated deterministic inversions with different regularization constraints. Features that appear in all models are interpreted as being well resolved by the data. Jackson (1976) and Meju & Hutton (1992) constructed extremal models that fit the data up to a given data misfit threshold with a most-squares inversion. This approach derives the extremal deviations of each model parameter from a best-fitting model. Kalscheuer & Pedersen (2007) used truncated singular value decomposition (TSVD) to estimate the model parameter errors and resolution of models from radio magnetotelluric (RMT) data. Finally, Kalscheuer *et al.* (2010) used the same approach to compare the errors and resolution properties of the RMT data against those of a joint inversion with electrical resistivity tomography data (ERT) and ERT data alone. The aforementioned methods partly account for model nonlinearity but violate formal

Bayesian principles, firstly because the "best" model is found by minimizing an objective function rather than analyzing the variables' marginal pdfs, and secondly because the estimated uncertainties are dependent on this best model, which in turn depends on the initial model used to find it (e.g., Chen *et al.* 2008). This poses questions regarding the statistical validity of the estimated model and parameter uncertainty.

The purpose of the present paper is to investigate MCMC-derived parameter uncertainty and bias of a finely parameterized 2D subsurface system for an increasing level of model constraints. In particular, we study how the posterior uncertainty changes when RMT data is inverted using (1) no constraints on the model structure, (2) smoothness constraints with different model norms and (3) joint inversion with ERT data. We also investigate the ability of the MCMC algorithm to retrieve the "true" measurement data errors and the regularization weight that provides appropriate weights to the model constraints.

The remainder of the paper is organized as follows: Section 2 presents the theoretical background of the proposed inversion approach. This is followed in Section 3 by the results of a synthetic model using different levels of model constraints and in Section 4 for a real world application using experimental data from an aquifer in Sweden. Section 5 discusses the implications of our results and highlights potential further developments. Finally, Section 6 concludes this paper with a summary of the presented work.

## 2. Method

### *2.1. Bayesian inversion*

Let the physical system under investigation be described by a vector of $M$ model parameters, $\mathbf{m} = (m_1, m_2, \ldots, m_M)$ and a set of $N$ observations, $\mathbf{d} = (d_1,d_2,\ldots,d_N)$ which are theoretically related to the model via a set of equations,

$$\mathbf{d} = g(\mathbf{m}) + \mathbf{e}, \tag{1}$$

where $\mathbf{e}$ is a vector of dimension $N$, which contains measurement data errors and any discrepancies caused by the model parameterization, deficiencies in the forward function $g(\mathbf{m})$, etc. The posterior pdf $p(\mathbf{m}|\mathbf{d})$ of the model parameters, conditional on the data, can be obtained by applying Bayes theorem (Tarantola & Valette 1982):

$$p(\mathbf{m}|\mathbf{d}) = \frac{p(\mathbf{m})p(\mathbf{d}|\mathbf{m})}{p(\mathbf{d})}, \tag{2}$$

where $p(\mathbf{d}|\mathbf{m})$ is the pdf of **d** conditional on **m**, also called the likelihood function $L(\mathbf{m}|\mathbf{d})$, $p(\mathbf{m})$ is the prior pdf and $p(\mathbf{d})$ signifies the evidence. The evidence is a normalizing constant that is required for Bayesian model selection and averaging (e.g., Malinverno 2002), but because our interests concern a fixed model parameterization, $p(\mathbf{d})$ can be removed without harm from eq. (2) leaving us with the following proportionality equality

$$p(\mathbf{m}|\mathbf{d}) \propto p(\mathbf{m})L(\mathbf{m}|\mathbf{d}). \tag{3}$$

The prior probability of the model vector, $p(\mathbf{m})$, represents the information known about the subsurface before collecting the actual data. It can be based on other types of geophysical measurements, geological information about the model structure, expected type of rocks and values of model parameters, etc. In the absence of detailed prior information about the subsurface properties, we assume a Jeffreys prior, that is, that the logarithm of each respective property is uniformly distributed (Jeffreys 1939; Tarantola 2005).

*2.2. The likelihood function*

The likelihood function summarizes the distance (typically a norm of a vector of residuals) between the model simulation and observed data. The larger the value of the likelihood, the closer the model response typically is to the experimental data. Under the assumption that the measurement data errors follow a normal distribution with zero mean, the likelihood function is given by (Tarantola 2005)

$$L(\mathbf{m}|\mathbf{d}) = \frac{1}{(2\pi)^{N/2}\det(\Sigma)^{1/2}}\exp\left(-\frac{1}{2}\left(g(\mathbf{m})-\mathbf{d}\right)^{\mathrm{T}}\Sigma^{-1}\left(g(\mathbf{m})-\mathbf{d}\right)\right), \tag{4}$$

where $\Sigma$ is the data covariance matrix and $\det(\Sigma)$ denotes the determinant of $\Sigma$. If the errors are uncorrelated, then $\Sigma$ is a diagonal matrix and $\det(\Sigma) = \prod_{i=1}^{N}\sigma_i^2$. The log-likelihood can then be expressed as

$$l(\mathbf{m}|\mathbf{d}) = -\frac{N}{2}\log(2\pi) - \frac{1}{2}\log(\prod_{i=1}^{N}\sigma_i^2) - \frac{1}{2}\phi_{d,2}, \quad (5)$$

where $\phi_{d,2} = \sum_{i=1}^{N}\left(\frac{g_i(\mathbf{m}) - d_i}{\sigma_i}\right)^2$ represents the data misfit and $\sigma_i$ denotes the standard deviation of the $i$-th measurement error. This misfit function is a measure of the distance between the forward response of the proposed model and the measured data, where the subscript defines the $l_2$ norm. The first term in eq. (5) is a constant, and the measurement data errors can be assumed unknown and estimated jointly with the model parameters. This approach is also referred to as hierarchical Bayes (e.g., Malinverno & Briggs 2004; Guo *et al.* 2011). As the data misfit becomes smaller, the log-likelihood increases and the proposed model is more likely to be a realization from the posterior distribution. Given the assumptions of the data errors made thus far, the sum of squared errors should follow a chi-square distribution with expected value of $N$. To avoid data over- or underfitting, it is therefore necessary to have a posterior misfit pdf with the same expected value.

When the data errors deviate from normality, it is common to use an exponential distribution, which is consistent with an $l_1$ norm instead of an $l_2$ norm (Menke 1998). Different publications have demonstrated that the $l_1$ norm is more robust against outliers, and often more realistic (e.g. Farquharson 1998; Shearer 1997). When the measurement errors are independent, the corresponding exponential likelihood function is given by (Tarantola 2005):

$$L(\mathbf{m}|\mathbf{d}) = \frac{1}{2^N\prod_{i=1}^{N}\sigma_i}\exp\left(-\sum_{i=1}^{N}\left|\frac{g_i(\mathbf{m}) - d_i}{\sigma_{\mathrm{d},i}}\right|\right), \quad (6)$$

which corresponds to the following formulation of the log-likelihood function

$$l(\mathbf{m}|\mathbf{d}) = -N\log(2) - \log(\prod_{i=1}^{N}\sigma_i) - \phi_{d,1}, \tag{7}$$

where the data misfit is now defined as $\phi_{d,1} = \sum_{i=1}^{N}\left|\frac{g_i(\mathbf{m}) - d_i}{\sigma_i}\right|$. This distribution has much longer tails (e.g., Menke 1989), thereby reducing the importance of outliers during parameter estimation.

*2.3. Constraining the model structure*

When strong a-priori knowledge of a suitable model structure is lacking, one may invert for the model pdf by only providing each model parameter's likely range of variation as *a-priori* information. An alternative is to also constrain the model structure to favor smooth spatial transitions. This is a common strategy in deterministic inversion (e.g. Constable *et al.* 1987; deGroot-Hedlin & Constable 1990), where these constraints serve as a regularization term that decreases the ill-posedness of the inverse problem. In the Bayesian framework, the constraints can be included in the prior pdf (e.g. Besag *et al.* 1995; Chen *et al.* 2012).

To favor models with smoothly varying resistivity structures, we impose independent normal distributions to the horizontal and vertical model gradients. This results in the following constraint prior pdf (see Appendix A)

$$c_{\mathrm{m},2}(\mathbf{m}) = \frac{1}{\left(2\pi\alpha_y^2\right)^{M_y}}\frac{1}{\left(2\pi\alpha_z^2\right)^{M_z}}\exp\left[-\frac{1}{2}\left(\frac{1}{\alpha_y^2}\mathbf{m}^{\mathrm{T}}\mathbf{D}_y^{\mathrm{T}}\mathbf{D}_y\mathbf{m} + \frac{1}{\alpha_z^2}\mathbf{m}^{\mathrm{T}}\mathbf{D}_z^{\mathrm{T}}\mathbf{D}_z\mathbf{m}\right)\right], \tag{8}$$

where $\mathbf{D}_y$ and $\mathbf{D}_z$ signify the difference operators in the horizontal and vertical directions with rank $M_y$ and $M_z$, respectively, $(M_y + 1)$ and $(M_z + 1)$ denote the number of horizontal and vertical grid cells, respectively, and $\alpha_y$ and $\alpha_z$ are the standard deviations of the model gradients in each spatial direction. If their expected values are similar for both directions, the constraint function becomes

$$\log(c_{m,2}(\mathbf{m})) = -(M_y + M_z)\log(2\pi\lambda^2) - \frac{1}{2}\phi_{m,2}, \tag{9}$$

where $\phi_{m,2} = \frac{1}{\lambda^2}\left(\mathbf{m}^{\mathrm{T}}\mathbf{D}_y^{\mathrm{T}}\mathbf{D}_y\mathbf{m} + \mathbf{m}^{\mathrm{T}}\mathbf{D}_z^{\mathrm{T}}\mathbf{D}_z\mathbf{m}\right)$ and $\lambda = \alpha_z = \alpha_y$ is a hyper-parameter to be determined using MCMC simulation. This latter variable bears much resemblance with model regularization weights used in deterministic inversions, and hence will be referred to as such hereafter. Note also that the right-hand side term in eq. (9) is essentially the model regularization term proposed by deGroot-Hedlin & Constable (2000). The smaller the value of $\lambda$, the higher the weight given to the regularization term.

Sharper spatial model transitions than those obtained by the least-squares smoothness constraints may be sought. In classical deterministic inversions, sharp transitions are usually imposed by applying alternative model norms (e.g., Farquharson 2008; Rosas Carbajal *et al.* 2012). Similar to how an exponential pdf was used to obtain more robust data misfit measures, here we apply it to increase the likelihood of models whose properties change abruptly from one cell to the next:

$$c_{m,1}(\mathbf{m}) = \frac{1}{\left(2\alpha_y\right)^{M_y}}\frac{1}{\left(2\alpha_y\right)^{M_z}}\exp\left[-\left(\frac{\left\|\mathbf{D}_y\mathbf{m}\right\|_1}{\alpha_y} + \frac{\left\|\mathbf{D}_z\mathbf{m}\right\|_1}{\alpha_z}\right)\right], \tag{10}$$

where a $l_1$ norm is used (subscript) for the smoothness constraints. In the case that $\alpha_z = \alpha_y = \lambda$, the log-distribution of eq. (10) becomes

$$\log(c_{m,1}(\mathbf{m})) = -(M_y + M_z)\log(2\lambda) - \frac{1}{\lambda}\left(\left\|\mathbf{D}_y\mathbf{m}\right\|_1 + \left\|\mathbf{D}_z\mathbf{m}\right\|_1\right). \tag{11}$$

The $l_1$ norm linearly weights the differences of the properties of adjacent cells. This is different from an $l_2$ norm that squares these differences, and hence an $l_1$ norm is less sensitive to sharp transitions between neighboring cells.

### *2.4 Forward computations*

To compute the likelihood functions described in the previous section, a numerical solver is needed to simulate the geophysical response of each proposed model. For both geophysical methods considered herein, the RMT and ERT responses are described by Maxwell's equations. In the general case, the model parameters and electromagnetic field vary dynamically in a 3D space. The higher the resolution of the resolved spatial dimension and the larger the number of model parameters, the more demanding the forward problem. Despite significant advances in computational power, 3D MCMC inversion remains a daunting computational task. We therefore focus our attention on a 2D model of the subsurface and compute the 2.5D ERT and RMT forward responses using finite-difference approximation. A detailed description of the forward solvers can be found in Kalscheuer *et al.* (2010), and interested readers are referred to this publication for additional details about the numerical setup and solution.

*2.5 Markov chain Monte Carlo strategy for high-dimensional problems*

For high-dimensional and non-linear inverse problems, it is practically impossible to analytically derive the posterior distribution. We therefore resort to MCMC sampling methods that iteratively search the space of feasible solutions. In short, MCMC simulation proceeds as follows. An initial starting point, $\mathbf{m}_{\text{old}}$ is drawn randomly by sampling from the prior distribution. The posterior density of this point is calculated by evaluating the product of the likelihood of the corresponding simulation and prior density. A new (candidate) point, $\mathbf{m}_{\text{new}}$ is subsequently created from a proposal distribution that is centered around the current point. This proposal is accepted with probability (Mosegaard & Tarantola 1995):

$$P_{\text{accept}} = \min\left\{1, \exp\left[l(\mathbf{m}_{\text{new}}|\mathbf{d}) - l(\mathbf{m}_{\text{old}}|\mathbf{d})\right]\right\} \qquad (12)$$

If the proposal is accepted the Markov chain moves to $\mathbf{m}_{\text{new}}$, otherwise the chain remains at its old location. After many iterations, the samples that are generated with this approach are distributed according to the underlying posterior distribution. The efficiency of sampling is strongly determined by the scale and orientation of the proposal distribution. If this distribution is incorrectly chosen, then the acceptance rate of candidate points might be unacceptably low, resulting in a very poor efficiency. On the contrary, if the proposal distribution is chosen accurately, the MCMC sampler will rapidly explore the posterior target distribution.

In this work, we use the MT-DREAM$_{(ZS)}$ algorithm (Laloy & Vrugt 2012), which was especially designed to efficiently explore high-dimensional posterior distributions. This is an adaptive MCMC algorithm (e.g., Roberts & Rosenthal, 2007), which runs multiple chains in parallel and combines multi-try sampling (Liu *et al.* 2000) with sampling from an archive of past states (Vrugt *et al.* 2009, see also Vrugt *et al.* 2008a) to accelerate convergence to a limiting distribution. Furthermore, it is fully parallelized and especially designed to run on a computer cluster. The MT-DREAM$_{(ZS)}$ algorithm satisfies detailed balance and ergodicity, and is generally superior to existing MCMC algorithms (Laloy & Vrugt 2012). To assess convergence, the Gelman-Rubin statistic (Gelman & Rubin 1992) is periodically computed using the last 50% of the samples in each of the chains. Convergence to a limiting distribution is declared if the Gelman-Rubin statistic is less than 1.2 for all parameters. After convergence, we use the last 25 percent of the samples in each chain to summarize the posterior distribution.

*2.6 Uncertainty estimation with most-squares inversion*

Most-squares inversion (Jackson 1976; Meju & Hutton 1992) is a deterministic inversion approach where extremal models are sought that fit the data up to a given threshold. First, a best fitting model $\mathbf{m}_0$ is calculated. Next, a particular cell of the model is chosen and the most-squares inversion is used to find the extremal values of this cell that satisfy a data misfit threshold $\phi_{d,2}^t = \phi_{d,2}[\mathbf{m}_0] + \Delta\phi$. All model cells are allowed to vary and two different searches are initiated to derive the smallest and largest acceptable resistivities. If we choose $\Delta\phi = 1$ it can be shown that this results in extremal values that deviate one standard deviation from the best fitting model (e.g., Kalscheuer *et al.* 2010). Most-squares inversion has been used to test the validity of other non-linear yet deterministic variance estimates, such as inversion schemes based on singular value decomposition (Kalscheuer & Pedersen 2007). Furthermore, it can also be applied with regularization constraints using the same model regularization weight used to derive the best fitting model and modifying the threshold misfit to $\phi_{d,2}^t = \phi_{d,2}[\mathbf{m}_0] + (1/\lambda^2)\phi_{m,2} + \Delta\phi$. The mean and uncertainty of the different cells derived from the most-squares inversion results are compared against their estimates from MCMC simulation.

## 3. Synthetic examples

To evaluate the impact of the model constraints and data on the posterior pdf, we consider a synthetic 2D resistivity model. This study is similar to the one presented by Kalscheuer *et al.* (2010). Two resistors and two conductors with thicknesses of 10 m (Fig. 1a) are immersed in a homogeneous medium of 100 Ωm. A conductor of 10 Ωm and 50 m length overlays a 1000 Ωm and 30 m long resistor at symmetric positions, and a resistor of 1000 Ωm and 50 m length overlays a 10 Ωm and 30 m long conductor, respectively. The transverse electric (TE) and transverse magnetic (TM) mode responses of this configuration were computed for the 17 different stations shown in Fig. 1(a). A total of 8 frequencies, regularly spaced on a logarithmic scale in the frequency range of 22 to 226 kHz were used, which resulted in a total of 544 data points. These synthetic observations were subsequently corrupted with a Gaussian measurement data error with standard deviation equal to three percent of the simulated impedances. To explicitly investigate the effect of the probabilistic properties of the measurement data errors, we also created a second data set by perturbing the error-free simulated forward responses with a zero-mean exponential distribution and a similar mean deviation of 3 percent of the modeled impedances. Unless stated differently, we refer to the RMT data as the data set contaminated with Gaussian noise in the remainder of this paper. To generate the synthetic ERT data, forward and reverse pole-dipole configurations were considered with electrodes placed at the positions of the 17 different RMT stations. Similarly to Kalscheuer *et al.* (2010), four expansion factors (1, 2, 4 and 6) and a basic potential electrode distance of 10 m, and level values of $n = 1,...,7$ for a fixed potential electrode distance were used. This resulted in a data set consisting of 306 different artificial observations. To mimic the effect of measurement data errors, the simulated data were again perturbed with a Gaussian error using a standard deviation of 3 percent of the simulated apparent resistivities. The model discretization used in the MCMC inversions is shown in Fig. 1(a). Each cell has dimensions of 5 × 10 m, but the cells located at the left, right and bottom edges of the domain extend until "infinity" (i.e., to accommodate the imposed boundary conditions). This results in a total of 228 different resistivity values that need to be estimated from the experimental data.

Figure 1(b) plots the final model derived from the RMT data using a classical deterministic inversion with smoothness constraints (c.f., deGroot-Hedlin & Constable 1990). This model was obtained after 3 iterations and has a misfit of $\phi_{d,2}$ = 533, assuming a 3 percent error of the impedance values. A homogenous half-space of 100 Ωm was used as the starting model. The inversion successfully retrieves the two shallow blocks, and indicates the presence of the deep conductor. However, it shows no evidence of the deep resistor. The resistivity value of the shallow conductor is well defined, but the magnitude of the resistor is underdetermined.

We now summarize the results of MCMC simulation using the different penalties of the model structure described previously in Section 2. Following recommendations made by Laloy & Vrugt (2012), we use three different chains and simultaneously create and evaluate five candidate points in each individual chain. To maximize computational efficiency, we run MT-DREAM$_{(ZS)}$ in parallel using sixteen different processors. Fifteen processors are used to simultaneously evaluate the different proposals, and achieve a linear speed up, whereas the remaining processor serves to execute the main algorithmic tasks of MT-DREAM$_{(ZS)}$. We invert for the log-resistivity values, and use a Jeffreys prior in the range of $10^{0.5}$ to $10^{3.5}$ Ωm. We also invert for the hyper-parameter $r$, which represents the standard deviation of the measurement data errors as a percentage of the measured impedances. We use a Jeffreys prior for $r$ as well, and define its upper and lower bound as half and double its true value (i.e., 1.5 to 6%). Appendix B details the log-likelihood that is used to estimate $r$ from the RMT data.

In the first MCMC trial, no constraints on the model structure (see eq. (5)) were specified. Convergence of the chains was reached after about 100,000 computational time units (*CTUs,* c.f. Laloy & Vrugt 2012a). Note that a single update of each of the parallel chains requires two *CTUs*, one for the evaluation of the candidate points, and one for the calculation of the posterior density of the reference set. To provide insights into the properties of the posterior resistivity distribution, Figure 2 displays four randomly chosen posterior models. The corresponding data misfit is also listed. The models exhibit an extreme variability and the only structure that is clearly persistent in all four realizations is the shallow conductor. Figures 3(a)-(c) depict ranges of the marginal posterior pdf of the resistivity of three vertical profiles. As expected, these results illustrate that model variability increases with depth. The first 20 meters appear rather well constrained by the data, but the uncertainty of the resistivity significantly increases beyond this depth. The data misfit and marginal posterior pdfs of the impedance error are represented with histograms in Figs. 3(d)-(e), respectively. The marginal distribution of the data misfit is centered on its a priori expected value of *N*, a finding that inspires confidence in the ability of MT-DREAM$_{(ZS)}$ to converge to the adequate parameter values. In other words, the proposed models do not systematically over or under fit the calibration data. Note also that the standard deviation of the relative data error is well resolved with mean value of $r = 0.03$ and standard deviation of 0.001 (see Fig. 3e).

To determine whether model constraints about the considered subsurface influence the efficiency and robustness of MCMC simulation, a second inversion was performed in which smoothly varying resistivity structures were favored by including Eq. (9) in the prior pdf. The prior distribution in this case is then the same Jeffreys distribution as before with the same parameter ranges, but multiplied by the exponential of Eq. (9). The regularization weight, $\lambda$ was assumed to follow a Jeffreys prior with range of half and two times the optimal value derived by fitting a normal distribution (eq. A2) to the true log-resistivity model. For convenience, we further assumed a similar value of $\lambda$ in both the vertical and horizontal direction.

Numerical results show that convergence was achieved after approximately 75,000 *CTUs*. Figure 4 illustrates that the posterior realizations exhibit far less spatial variability than those previously derived for the unconstrained case without smoothness constraints, although the models are visually quite different. This is further confirmed by the vertical resistivity profiles depicted in Figs. 5a-c. Model parameter uncertainty has significantly reduced, but with the side effect that some features of the true model are no longer accurately represented in the posterior pdf. Indeed, the two conductors and the shallow resistor are clearly detected, but the deep resistor is not adequately resolved. Yet, the MCMC inferred resistivity increases with depth, which is consistent with the observations. The marginal distribution of the data misfit presented in Fig. 5(d) again nicely centers on the true value, and is quite similar to the unconstrained inversion trial. The same is true for the data error estimation (Fig. 5f): the true value is obtained and the variability is similar to that previously observed in Fig. 3(e). The estimated value of $\lambda$ is slightly larger than its previous counterpart derived from the true log-resistivity model. This finding is to be expected and is a direct consequence of the influence of the data misfit term in the estimation (i.e., less weight is put on the model constraints).

We now summarize the MCMC results with an $l_1$ measure (see eq. 11) for the model constraints. For this inversion, we use a data set contaminated with exponentially distributed errors and log-likelihood function given by Eq. (7). For consistency, we again use a Jeffreys prior for all regular model parameters (resistivities) and hyper-parameters (regularization weight and impedance error). The resistivity and impedance error prior bounds remain the same as in the past examples, but the prior of the regularization weight ranges from half (0.055) to four (0.44) times the value found by fitting Eq. (11) to the true resistivity model. We purposely increased the upper bound of $\lambda$ so that the posterior pdf was unaffected by the a priori bounds.

About 67,000 *CTUs* were needed to declare convergence to a limiting distribution. The posterior realizations presented in Fig. 6 are rather homogeneous, and display even less variability than their counterparts previously depicted in Fig. 4 using the least-squares model constraints. The two shallow features are clearly identified, and a deep conductor can be seen in three of the four figures. The deep resistor however is not evident in any of the models. This becomes more evident if we plot the three depth profiles (Figs 7a-c). The 95% posterior uncertainty ranges are comparable to those obtained with the inversion using the $l_2$ model constraints. The data misfit and the impedance errors are very well recovered. However, the posterior mean of $\lambda$ is substantially larger than its value derived from fitting the true model structure to an exponential model (0.11).

Finally, we jointly invert the RMT and ERT data using least-squares smoothness constraints. In this particular case, the log-likelihood function is given by the sum of those corresponding to each data set. A derivation of the ERT likelihood is presented in Appendix C. This inversion includes the ERT data error, which constitutes a new hyper-parameter to be estimated. We use a Jeffreys prior for this parameter, with bounds given by half and twice its true value.

Convergence of the chains was achieved after about 60,000 *CTUs*. The posterior realizations shown in Fig. 8 clearly resolve the two conductors and the two resistors. The vertical resistivity profiles presented in Figs. 9a-c confirm that joint inversion improves parameter convergence. Yet, the resistor below the conductor (Fig. 9a) is not particularly well resolved. But, its magnitude is much better estimated than in the previous inversions. The model constraints enforce smooth transitions from the conductor to the resistor and vice versa, which complicates estimation of the actual magnitudes in the vicinity of these transitions (e.g., Fig. 9c below the conductor). The posterior histograms of the RMT (Fig. 9d) and ERT data (Fig. 9e) misfits are closely centered on their true values, a desirable finding that indicates that both data types are equally important in the fitting of the parameters. The marginal posterior distribution of the regularization weight (Fig. 9f) demonstrates a tendency towards somewhat larger values than obtained from the RMT data. This is not surprising, as new data have been added to the likelihood function. For completeness, Figs. 9g-h plot histograms of the impedance and apparent resistivity error. The posterior ranges encompass the synthetic true values, although the most likely (expected) values are somewhat smaller. This demonstrates that the measurement errors of both data types can be successfully retrieved from the joint inversion presented herein.

To provide more insights into the behavior of the MT-DREAM$_{(ZS)}$ algorithm, Fig. 10 presents the evolution of the sampled model structure in one randomly chosen chain as a function of the number of MCMC realizations. The true value and those inferred from the different MCMC trials are given by the $l_2$ norm of the difference operator applied to the model vector in the horizontal and vertical directions (i.e., the term enclosed in parentheses in eq. (9)). We restrict our attention to the posterior samples – thus after burn-in (c.f. Laloy & Vrugt 2012a) has been achieved.

The MCMC inversion without model constrains (Fig. 10a) converges to a model structure that overestimates the actual variability observed in the true model. The true model is not contained in the sampled posterior pdf. When smoothness constraints are explicitly included in the formulation of the log-likelihood function, the posterior models converge much closer to the true model, but with insufficient structure. This is particularly true if the $l_1$ norm is used. The average model structure in this case is 24, which is about half the true value. The correspondence between the true model and posterior realizations improves somewhat if an $l_2$ norm is used. Indeed, the sampled chain trajectory moves closer to the dashed black line, but nevertheless the actual model variability is still underestimated. Fortunately, a joint inversion of RMT and ERT data provides posterior realizations with properties similar to that of the true model, especially if an $l_2$ norm is used for the model constraints.

Table 1 lists the center values and standard deviations estimated with the MCMC and most-squares inversions for the cells shown in Fig 1(a). To enable a comparison between both methods, we calculate two different standard deviations from the posterior mean MCMC model: one for resistivity decrease and one for resistivity increase. We performed three most-squares inversions: one for the RMT data with smoothness constraints, one for the ERT data with smoothness constraints, and one for joint inversion with smoothness constraints. To find the best fitting models, we locate that sample of the MCMC chains with largest value of the sum of eqs. (5) and (9). This model was then used to initiate a deterministic inversion with additional Marquardt-Levenberg damping (cf. Kalscheuer *et al.* 2010) to attempt to find a model with an even larger summed log-likelihood. This model was then used by the most-squares inversion to find the extremal values of each cell. In both inversion steps, we used the mean model regularization weight determined by the MCMC inversions. As seen in Fig. 1(b), the model discretization is finer in the horizontal direction for the most-squares inversion. At each iteration we therefore averaged the two resistivities involved in each particular cell to force a single resistivity value and make it comparable to the MCMC inversion cell.

The standard deviations summarized in Table 1 show that the two types of inversions provide similar uncertainty estimates. However, the standard deviations derived with the most-squares inversion are consistently larger than those derived with MCMC simulation. For example, in the single inversions of the RMT data, cell B has standard deviations of 0.18/0.19 for the MCMC inversion, and 0.24/0.24 for the most-squares inversion, respectively. These differences appear larger for the joint inversion. For instance, cell A has standard deviations of 0.08/0.08 with the MCMC inversion, but with the most-squares inversion these values have doubled. Furthermore, we see that the mean value estimates are quite different for the two types of inversion. For example, the mean value of cell A for the ERT data and MCMC inversion is 1.0, whereas its counterpart derived from the most-squares inversion is 1.16. Thus, although the width of the uncertainty ranges can be quite similar, the mean value might induce shifts in the posterior distribution.

## 4. Field data example: Skediga Area (Sweden)

We now apply our methodology to real-world RMT data. A tensor RMT survey was conducted in Skediga (Sweden) to determine the geometry of a glacio-fluvial aquifer system composed of a sand/gravel formation overlying crystalline basement. The aquifer system is overlain by a formation dominated by clay lenses. We use the same RMT data as Kalscheuer & Pedersen (2007), that is, 528 data points consisting of apparent resistivities and phases of the determinant mode (Pedersen & Engels 2005), acquired at 22 different stations using 12 frequencies in the range of 4 to 181 kHz. An estimate of the data error was provided by the impedance estimation from the electric and magnetic field measurements and an error floor of 1.5% was used as in the previous studies (Pedersen *et al.* 2005, Kalscheuer & Pedersen 2007). The error floor constitutes a lower bound to the estimated data errors such that no single data has an error estimate smaller than this value.

Figure 11(a) shows the model obtained by Kalscheuer & Pedersen (2007) derived from a deterministic inversion with smoothness constraints using a half-space of 1000 Ωm as the initial model. The model was obtained after four iterations and has a data misfit of $\phi_{d,2} = 1141$. Pedersen *et al.* (2005) interpret the 30 Ωm iso-line (i.e. the transition between the two greenish colors) as the lower bound of the clay lenses. According to boreholes in the vicinity of the profiles, the transition from the aquifer to the underlying crystalline basement occurs at about 30 m depth (Kalscheuer & Pedersen 2007).

We ran the MT-DREAM$_{(ZS)}$ algorithm on a 2D domain consisting of 288 model parameters using the $l_2$ smoothness constraints. Each resistivity cell is of size 5 × 10 m, except for the edges that extend to the end of the forward mesh (1300 m in each direction). We used Jeffreys priors in the range of $10^{0.5}$ to $10^{3.5}$ of $\rho\,(\Omega\text{m})$. In addition, we estimated two hyper-parameters: the regularization weight $\lambda$ and a data error correction factor. The latter represents a scaling factor of the errors and error floor. We assume a Jeffreys prior for this scaling factor, with ranges between the logarithms of 0.5 and 4.

Convergence was reached after approximately 150,000 *CTUs*. Figures 11(b) and (c) show two realizations from the MCMC derived posterior pdf. The two models clearly indicate two shallow conductors at profile offsets of 40 m and between 170 m and 220 m. A deep resistor is also found that is deeper on the left side of the profile than in the middle and that disappears on the right side. A mean posterior model was constructed by taking the mean value of the different realizations of the posterior pdf (Fig. 11d). This model is largely comparable to the model obtained by the deterministic inversion; the clay – sand/gravel transitions are located at similar depths nearly everywhere along the profile and the overall basement geometry of the two different models corresponds well (this was also noted with the ensemble mean of the synthetic example using least squares smoothness constraints compared to Fig. 1b, not shown here). Some deviations are possibly due to difference in model discretization, but may more probably be due to differences in data fitting, as discussed below.

We present four vertical profiles of the posterior pdf in Figs. 12(a)-(d), at offsets (a) $y$ = 50 m, (b) $y$ = 100 m, (c) $y$ = 150 m and (d) $y$ = 200 m. As expected, the profiles show an increase in model variability below the conductive clay lenses. Furthermore, we see how the clay – sand/gravel transitions are much better determined at places where the aquifer stretches up to the surface (Figs. 12b and c). In these regions there is no overlapping between the two resistivity intervals, whereas in the other two profiles the transition happens more smoothly, probably due to the model constraints. Also the transition to a fixed basement resistivity is smooth because of the model regularization. Magnitudes are expected to be above $\rho = 1000\ \Omega\text{m}$ for the crystalline basement (Pedersen *et al*. 2005). These values are reached at all profiles except in Fig. 12(d), probably due to the important clay thickness in the shallow part of the model. Figures 12(e)-(f) show marginal distributions of the posterior data misfit and the data error correction factor. These two variables are related. The mean data misfit is 542 and the number of data is comprised within the estimated data misfit uncertainty range. The mean data error correction factor is 1.84, hence data errors are estimated to be almost twice those initially assumed for the impedances. The data misfits presented in Fig. 11 are calculated using data errors corrected with this value, and they show that the model given by the deterministic inversion appears to be over-fitting the data. This, in turn, could explain the differences in magnitude observed between the two models. An inversion of the Skediga data set with the same priors for the error scaling factor and resistivity values but with no model constraints converged to a similar marginal posterior pdf of the impedance errors (not shown). In accordance with the synthetic example, the posterior pdf of the unconstrained inversion contains models with unrealistically high spatial variability.

## 5. Discussion

We have presented the first fully 2D pixel-based MCMC inversion of plane-wave EM data. While the presented results indicate that the inversion can be successfully addressed within a probabilistic framework, notable features and issues arise that are discussed in more detail below.

A comparison between the most-squares and MCMC inversions showed that while the former tends to provide slightly larger uncertainty estimates, the results of the two approaches are comparable. A more substantial difference between the methods relates to the center values from which the uncertainty estimates are derived. This difference is mainly caused by the fact that the most-squares inversion starts from a model that minimizes the combined data and model misfit function, while the MCMC analysis is based on an ensemble mean model obtained from a combination of the marginal estimates of individual variables. The minimization approach used in the most-squares inversion is not rigorously formal, as the best model should be the one that best represents the statistics of the posterior pdf rather than the minimization of the combined data and model misfit function. Calculating maximal and minimal perturbations of specific parameters from this "optimal" model could be the reason for the "shifted" and slightly larger uncertainty ranges compared to the MCMC estimates that describe the ensemble statistics of the posterior pdf.

The type of model parameterization and the number of parameters have an important impact on the posterior pdfs. Laloy *et al.* (2012) and Linde & Vrugt (2013) used model parameterizations based on Legendre polynomials and the discrete cosine transform, respectively, to show how improper model truncations may lead to biased model estimates. To alleviate this problem, we considered a finely discretized model. However, the unconstrained inversions converge to models that exhibit much more structure than the true model (see Fig. 10a), which is in agreement with Linde & Vrugt (2013). When running inversions with coarser grids (i.e., 10 × 10 m cells, not shown herein), the proposed models and the true model are in much better agreement and the uncertainty ranges of the parameters were strongly reduced. This highlights the fundamental trade-off between model resolution and variability: allowing a higher spatial resolution by using smaller model cells implies larger resistivity ranges for each pixel.

To obtain meaningful results for fine model discretizations, it appears fundamental to add additional constraints regarding the model structure. As noted by Grandis *et al.* (1999) for the 1D MT problem, the use of least-squares smoothness constraints reduced the presence of unrealistic oscillations in the models and led to smaller and more realistic estimates of parameter uncertainty. Unfortunately, the models provided by the constrained inversions did not contain all the features of the true model. In regions where the data are not sensitive enough, the model constraints strongly affect the resulting parameter values and result in biased estimates.

The problem of biased estimates was partly mitigated through joint inversion of the plane-wave EM data with ERT. The inversion of the ERT data alone with $l_2$ smoothness constraints (not shown) did recover the deep resistor albeit with a smaller magnitude than the true value, but not the deep conductor that was resolved by the RMT data. As seen in Fig. 10, when inverting the ERT data and plane-wave EM data separately, constraining the model structure led to oversimplified models, whereas the joint inversion led to the correct amount of model structure for this specific application. The models obtained from the plane-wave EM data could clearly be improved by adding lower frequencies, while a larger electrode spread would improve the ERT models. However, our intention was not to determine an optimal experimental design, but to evaluate the implications of the different constraints applied to the inferred subsurface models. In this sense, we see how the combination of two complimentary methods helps to better estimate the resistivity models in terms of structure and magnitude, and effectively reduces the weight given to the model constraints.

Other strategies can also be applied to tackle the aforementioned issues. The incorporation of a pre-supposed geostatistical model or summary statistics derived from training images can easily be incorporated in the Bayesian framework (e.g. Cordua *et al*., 2012). Clearly, the resulting models would be much closer to the true model if the true model structure was known and we penalized deviations from this value in eqs. (9) and (11), rather than penalizing deviations from zero variability. Reliable information of this kind is often not available and strong assumptions about the model structure will to a certain degree promulgate biased model estimates. Nevertheless, it might be favorable to test the resulting models under such restrictive assumptions, rather than to obtain models that are too variable to be meaningful.

Alternatively, one may consider a set of possible model parameterizations, model discretizations and/or model constraints that may seem equally suitable for a specific problem. In the spirit of Oldenburg & Li (1991), one may test the different hypotheses of the model structure and compare the results. More quantitatively, a 2D trans-dimensional inversion algorithm could be implemented. The trans-dimenional algorithm would, for a chosen parameterization, estimate the appropriate degree of discretization, while inherently favoring models with fewer parameters (see Bodin & Sambridge 2009 for a 2D application to seismic tomography). The implementation of such a method is beyond the scope of the present work. Possibly more interesting than determining appropriate model discretizations would be to determine preferred model parameterizations. In fact, a formal theory based on Bayes factors (e.g. Kass & Raftery 1995) could be used to evaluate evidence in favor of a null hypothesis (see Khan & Mosegaard 2002 and Khan *et al.* 2004 for applications of Bayes factors to study the physical properties of the moon). Bayes factors could be used within a model selection strategy to evaluate the a posteriori probability of different model parameterizations and discretizations. We leave such a study of Bayesian hypothesis testing for future work.

## 6. Conclusions

We presented the first pixel-based and fully 2D MCMC inversion of plane-wave EM and ERT data. The results of the inversion include the posterior mean and uncertainty of the model parameter estimates. Numerical findings demonstrated a necessity to add explicit constraints on the model structure to obtain meaningful results. These constraints were designed such that they favor model parsimony, and consequently the posterior ensemble mean was shifted closer to that of its true value. However, model interpretation should be done with some care, acknowledging that models may be biased in regions with insufficient data sensitivity, and uncertainty estimates are determined by the imposed model constraints.

The MCMC inversion not only appropriately converged to the posterior mean model, the posterior realizations adequately estimated the actual data errors, including a regularization weight that favors the appropriate model structure. Joint inversion of the ERT and plane-wave EM data provided the best model estimates. The inversion methodology was applied to real RMT aquifer data from Sweden. The MCMC derived posterior mean model was very similar to that of the model geometry obtained from a deterministic inversion. On top of this, the MT-DREAM$_{(ZS)}$ algorithm also retrieved a correction of the impedance errors, which suggested that the deterministic inversion might have over-fitted the experimental data. The differences among the resistivity magnitudes of the two different models may hence be explained by a difference in data fitting. Future work should involve diagnostic criteria and methodologies that help favor model selection. In this regard, Bayes factors may be of particular interest.

**Acknowledgments**

We thank Jinsong Chen, Amir Khan and the editor Mark Everett for their very helpful comments that improved the quality of the paper. Laust B. Pedersen, from Uppsala University, kindly provided the RMT data from Skediga, Sweden. The source code of the MT-DREAM$_{(ZS)}$ algorithm can be obtained from the last author upon request. This research was supported by the Swiss National Science Foundation under grant 200021-130200.

**Table and Figure captions**

**Table 1.** Mean values and standard deviations of the cells highlighted in Fig. 1(a) for individual and joint MCMC and most-squares (MS) inversions with different types of model constraints. The center values are the mean values for the MCMC inversions and the parameter derived from the best-fitting MCMC model for the most-squares inversions (cf. section 3 for detail). The standard deviations (SD) are given in logarithmic units that are calculated individually for each side of the center value (+/).

**Figure 1.** (a) Synthetic test model with the MCMC model discretization highlighted. Letters A, B, C, and D indicate cells for which the inversion results are evaluated against those of deterministic most-squares inversions. Numbered letters V1, V2 and V3 indicate the offsets at which the resistivity marginal posterior pdfs are presented. (b) Model obtained by inverting RMT data (3% error on the impedance elements) with a smoothness constrained deterministic inversion. The mesh in (b) corresponds to the model discretization of the deterministic inversions and the forward modeling mesh. The triangles at the top of the figures indicate the locations of the RMT stations and the ERT electrodes.

**Figure 2.** (a-d) Posterior MCMC realizations from the inversion of RMT data with no model constraints other than minimal and maximal parameter bounds of $\rho = 10^{0.5}$ and $10^{3.5}$ Ωm, respectively. It is very difficult to identify a clear correlation between these realizations and the true underlying model in Fig. 1(a).

**Figure 3.** MCMC inversion of RMT data without model constrains. (a–c) Marginal posterior pdf of the vertical profiles V1, V2 and V3 corresponding to the offsets (a) 55 m, (b) 95 m, and (c) 135 m. The red line represents the true values, while the solid and dashed blue lines represent the mean and P2.5 and P97.5 percentiles, respectively. It is seen that below ~30 m the posterior models span the full prior range of resistivity. Grey color-coding indicates the full posterior pdf range. Histograms of the (d) data misfit and (e) the inferred impedance error marginal posterior pdf. The red crosses at the top of the histograms depict the values corresponding to (d) the data misfit of the true model and (e) the true error standard deviation.

**Figure 4.** (a-d) Posterior MCMC realizations obtained by inverting the RMT data with least-squares smoothness constrains. All the four anomalous bodies are somewhat indicated, even if it is only the upper left conductive body that is well resolved.

**Figure 5.** MCMC inversion of RMT data with least-squares smoothness constrains. (a–c) Marginal posterior pdfs of the vertical profiles V1, V2 and V3 corresponding to the offsets (a) 55 m, (b) 95 m, and (c) 135 m. The red line represents the true values, while the solid and dashed blue lines represent the mean and P2.5 and P97.5 percentiles, respectively. Grey color-coding indicates the full posterior pdf range. It is clear that the smoothness constraints have largely decreased model variability. Histograms of the (d) data misfit, (e) regularization weight and (f) impedance error marginal posterior pdf. The red crosses at the top of the histograms depict (d) and (f) the true values and (e) the value given by fitting eq. (9) to the true log-resistivity model.

**Figure 6.** (a-d) Posterior MCMC realizations obtained by inverting the RMT data with $l_1$ smoothness constrains. The upper anomalous bodies are resolved, but not the lower ones.

**Figure 7.** MCMC inversion of RMT data with $l_1$ smoothness constrains. (a–c) Resistivity marginal posterior pdf of the vertical profiles V1, V2 and V3 corresponding to the offsets (a) 55 m, (b) 95 m, and (c) 135 m. The red line represents the true values, while the solid and dashed blue lines represent the mean and P2.5 and P97.5 percentiles, respectively. Grey color-coding indicates the full posterior pdf range. The parameters' uncertainties are comparable to those of the $l_2$ smoothness constrains. Histograms of the (d) data misfit, (e) regularization weight and (f) impedance error marginal posterior pdf. The red crosses at the top of the histograms of (d) and (f) depict the true values. (e) The value given by fitting eq. (11) to the true log-resistivity model (0.11) is not comprised in the marginal posterior pdf.

**Figure 8.** (a-d) Posterior MCMC realizations obtained by joint inversion of RMT and ERT data with least-squares smoothness constrains. The anomalous bodies are better defined compared with the inversions of RMT data alone (see Fig. 4).

**Figure 9.** MCMC joint inversion of RMT and ERT data with least-squares smoothness constrains. (a–c) Resistivity marginal posterior pdfs of the vertical profiles V1, V2 and V3 corresponding to the offsets (a) 55 m, (b) 95 m, and (c) 135 m. The red line represents the true values, while the solid and dashed blue lines represent the mean and P2.5 and P97.5 percentiles, respectively. Grey color-coding indicates the full posterior pdf range. The range of the posterior pdf is rather small, but covers essentially the true model. Histograms of the (d) RMT data misfit, (e) ERT data misfit, (f) regularization weight, (g) RMT impedance error and (h) ERT apparent resistivity error marginal posterior pdfs. The red crosses at the top of the histograms depict (d), (e), (g) and (h) the true values and (f) the value given by fitting eq. (9) to the true log-resistivity model.

**Figure 10.** Posterior least-squares model structure metric as a function of realization number for the different types of MCMC inversions considered. (a) MCMC inversion of RMT data without model constrains. This inversion needs many more realizations to converge than all other cases and has a much larger average model structure. (b) MCMC inversions with model constraints. The dashed black line represents the true value. The joint inversion of RMT and ERT is the only case that proposes models with the same amount of model structure as the true model.

**Figure 11.** (a) Deterministic inversion model obtained from RMT data acquired at Skediga, Sweden (modified after Kalscheuer and Pedersen (2007)). Numbered letters V1, V2, V3 and V4 indicate the offsets at which the resistivity marginal posterior pdfs are presented in Fig. 12. (b-c) Posterior MCMC realizations obtained by inversion of the same data with least-squares smoothness constrains. (d) Ensemble posterior mean model from MCMC inversion. The data misfits are calculated with errors inferred from the mean value of Fig. 12(e). Note the strong similarity between the models in (a) and (d).

**Figure 12.** MCMC inversion of the Skediga data set with least-squares model constraints. (a–d) Resistivity marginal posterior pdf of the vertical profiles V1, V2, V3 and V4 corresponding to the offsets (a) 50 m, (b) 100 m, (c) 150 m and (d) 200 m of the model shown in Fig. 11(d). The solid and dashed blue lines represent the mean and P2.5 and P97.5 percentiles, respectively. The red line represents the values obtained with the deterministic inversion (see Fig. 11a). Grey color-coding indicates the full posterior pdf range. (e-f) Histograms of the (e) data misfit and (f) impedance error scaling factor marginal posterior pdfs. The red cross at the top of (e) depicts the number of data.

**Appendix A: 2D Smoothness constraints**

To obtain smoothly varying model property variations in the 2D models, we impose zero-mean normal prior distributions with respect to the vertical and horizontal log-resistivity gradients:

$$\begin{cases} c_{m,2}^{y}(\mathbf{m}) = \dfrac{1}{\left(2\pi\alpha_y^2\right)^{M_y}} \exp\left[-\dfrac{1}{2\alpha_y^2}\left(\mathbf{m}^{\mathrm{T}}\mathbf{D}_y^{\mathrm{T}}\mathbf{D}_y\mathbf{m}\right)\right] \\ c_{m,2}^{z}(\mathbf{m}) = \dfrac{1}{\left(2\pi\alpha_z^2\right)^{M_z}} \exp\left[-\dfrac{1}{2\alpha_z^2}\left(\mathbf{m}^{\mathrm{T}}\mathbf{D}_z^{\mathrm{T}}\mathbf{D}_z\mathbf{m}\right)\right] \end{cases}, \quad \text{(A1)}$$

where $\mathbf{D}_y$ and $\mathbf{D}_z$ are the difference operators in the horizontal and vertical directions with rank $M_y$ and $M_z$ respectively, and $\alpha_y$ and $\alpha_z$ are the standard deviations of the log-resistivity gradients in each direction. Assuming that the two pdfs are uncorrelated, the joint pdf of the horizontal and vertical resistivity gradients is given by multiplication of each pdf (Eq. 8). When the standard deviations are the same, eq. (8) can be expressed as

$$c_{m,2}(\mathbf{m}) = \frac{1}{\left(2\pi\lambda^2\right)^{M_y}} \frac{1}{\left(2\pi\lambda^2\right)^{M_z}} \exp\left[-\frac{1}{2\lambda^2}\left(\mathbf{m}^{\mathrm{T}}\mathbf{D}_y^{\mathrm{T}}\mathbf{D}_y\mathbf{m} + \mathbf{m}^{\mathrm{T}}\mathbf{D}_z^{\mathrm{T}}\mathbf{D}_z\mathbf{m}\right)\right], \quad \text{(A2)}$$

where $\lambda = \alpha_z = \alpha_y$. Taking the logarithm of eq. (A2) results in

$$\log\left(c_{m,2}(\mathbf{m})\right) = -M_y \log\left(2\pi\lambda^2\right) - M_z \log\left(2\pi\lambda^2\right) - \frac{1}{2\lambda^2}\left(\mathbf{m}^{\mathrm{T}}\mathbf{D}_y^{\mathrm{T}}\mathbf{D}_y\mathbf{m} + \mathbf{m}^{\mathrm{T}}\mathbf{D}_z^{\mathrm{T}}\mathbf{D}_z\mathbf{m}\right), \quad \text{(A3)}$$

or, equivalently

$$\log\left(c_{m,2}(\mathbf{m})\right) = -\left(M_y + M_z\right)\log\left(2\pi\lambda^2\right) - \frac{1}{2\lambda^2}\left(\mathbf{m}^{\mathrm{T}}\mathbf{D}_y^{\mathrm{T}}\mathbf{D}_y\mathbf{m} + \mathbf{m}^{\mathrm{T}}\mathbf{D}_z^{\mathrm{T}}\mathbf{D}_z\mathbf{m}\right). \quad \text{(A4)}$$

**Appendix B: log-likelihood function for plane-wave EM data**

Equation (5) represents the log-likelihood function of a set of normally distributed errors that have zero mean and are uncorrelated. These errors may, however, have different standard deviations. Indeed, RMT data often comprise apparent resistivities and phases. Let the first $N/2$ data points be the apparent resistivities $d_i = \rho_i^{app}$, $i = 1,...,N/2$, and the last $N/2$ data points the phases $d_i = \phi_i$, $i = N/2+1,...,N$. The data standard deviations can then be expressed as (Fischer & LeQuang 1981)

$$\sigma_i = \begin{cases} rd_i, & \text{if } i = 1,...,N/2 \\ \dfrac{r}{2}, & \text{if } i = N/2+1,...,N \end{cases}, \tag{B1}$$

where $r$ is the standard deviation of the relative error of the apparent resistivities, which is assumed to be the same for all measurements. Using eq. (B1), the middle term in eq. (5) can be expressed as

$$\frac{1}{2}\log\left(\prod_{i=1}^{N}\sigma_i^2\right) = \frac{1}{2}\log\left(\prod_{i=1}^{N/2}\left(r\rho_i^{app}\right)^2 \prod_{i=N/2+1}^{N}\left(\frac{r}{2}\right)^2\right), \tag{B2}$$

which leads to,

$$\frac{1}{2}\log\left(\prod_{i=1}^{N}\sigma_i^2\right) = \log\left(\frac{r^N}{2^{N/2}}\prod_{i=1}^{N/2}\rho_i^{app}\right). \tag{B3}$$

Expanding the logarithm and replacing this expression in eq. (5) gives

$$l(\mathbf{m}|\mathbf{d}) = -\frac{N}{2}\log(2\pi) + \frac{N}{2}\log(2) - N\log(r) - \sum_{i=1}^{N/2}\log\rho_i^{app} - \frac{1}{2}\sum_{i=1}^{N}\left(\frac{g_i(\mathbf{m}) - d_i}{\sigma_i}\right)^2, \tag{B4}$$

which is equivalent to

$$l(\mathbf{m}|\mathbf{d}) = -\frac{N}{2}\log(\pi) - N\log(r) - \sum_{i=1}^{N/2}\log\rho_i^{app} - \frac{1}{2}\sum_{i=1}^{N}\left(\frac{g_i(\mathbf{m}) - d_i}{\sigma_i}\right)^2. \tag{B5}$$

**Appendix C: log-likelihood functions for ERT data**

In the case of ERT, we consider a single type of data. The apparent resistivities are assumed to comprise relative errors. Therefore, we follow the same derivation as in Appendix B, but with standard deviations given by $\sigma_i = rd_i,\ i = 1,...,N$ . Then, the middle term of eq. (5) can be expressed as

$$\frac{1}{2}\log\left(\prod_{i=1}^{N}\sigma_i^2\right) = \log\left(r^N\prod_{i=1}^{N}\rho_i^{app}\right), \tag{C1}$$

which leads to a log-likelihood of the form

$$l(\mathbf{m}|\mathbf{d}) = -\frac{N}{2}\log(2\pi) - N\log(r) - \sum_{i=1}^{N}\log\rho_i^{app} - \frac{1}{2}\sum_{i=1}^{N}\left(\frac{g_i(\mathbf{m}) - d_i}{\sigma_i}\right)^2. \tag{C2}$$

**Table 1.**

| Type of inversion | Model constraint | Cell A | | Cell B | | Cell C | | Cell D | |
|---|---|---|---|---|---|---|---|---|---|
| | | Center | SD (- / +) | Center | SD (- / +) | Center | SD (- / +) | Center | SD (- / +) |
| | | $\log_{10} \rho$ ($\Omega$m) | | $\log_{10} \rho$ ($\Omega$m) | | $\log_{10} \rho$ ($\Omega$m) | | $\log_{10} \rho$ ($\Omega$m) | |
| Individual RMT MCMC | $l_2$-difference | 0.97 | 0.12/ 0.11 | 2.04 | 0.18/ 0.19 | 2.36 | 0.11/ 0.15 | 1.36 | 0.21/ 0.21 |
| Individual RMT MS | $l_2$-difference | 0.98 | 0.15/ 0.12 | 1.90 | 0.24/ 0.24 | 2.36 | 0.19/ 0.17 | 1.09 | 0.22/ 0.26 |
| Individual ERT MCMC | $l_2$-difference | 1.00 | 0.10/ 0.09 | 2.00 | 0.12/ 0.10 | 2.65 | 0.11/ 0.11 | 2.05 | 0.14/ 0.14 |
| Individual ERT MS | $l_2$-difference | 1.16 | 0.17/ 0.17 | 1.63 | 0.23/ 0.24 | 2.64 | 0.18/ 0.18 | 2.12 | 0.23/ 0.23 |
| Joint MCMC | $l_2$-difference | 0.94 | 0.08/ 0.08 | 2.35 | 0.18/ 0.18 | 2.78 | 0.17/ 0.15 | 1.13 | 0.23/ 0.25 |
| Joint MS | $l_2$-difference | 0.99 | 0.15/ 0.16 | 2.18 | 0.25/ 0.25 | 3.11 | 0.20/ 0.18 | 1.05 | 0.22/ 0.26 |
| True values | - | 1.0 | N/A | 3.0 | N/A | 3.0 | N/A | 1.0 | N/A |

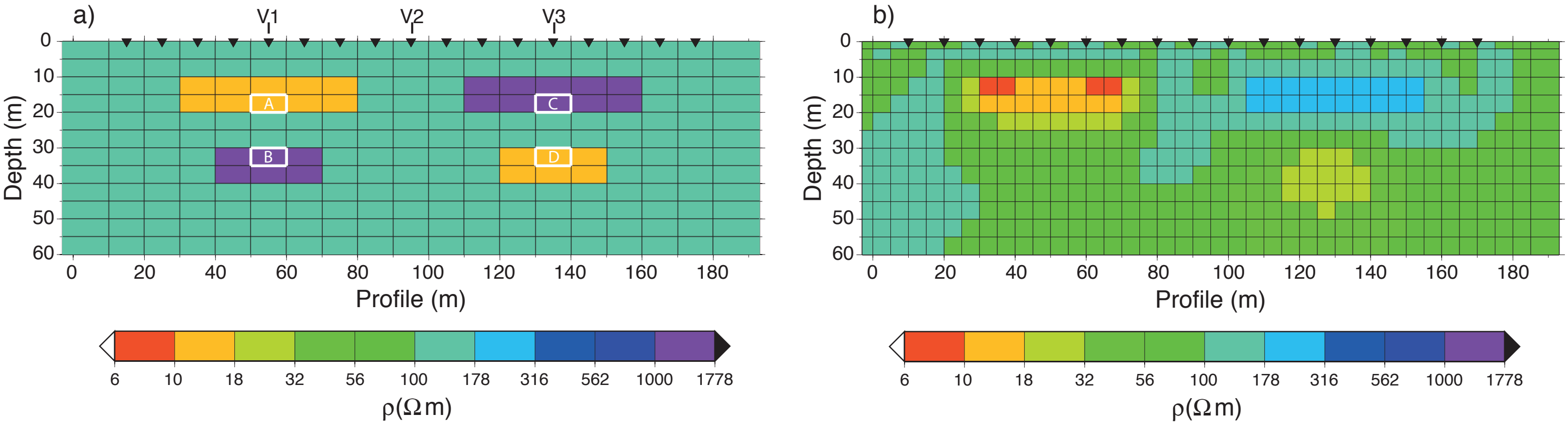

a)
V1
V2
V3
A
B
C
D
Depth (m)
Profile (m)
0 20 40 60 80 100 120 140 160 180
0 10 20 30 40 50 60
6 10 18 32 56 100 178 316 562 1000 1778
ρ(Ωm)
b)
Depth (m)
Profile (m)
0 20 40 60 80 100 120 140 160 180
0 10 20 30 40 50 60
6 10 18 32 56 100 178 316 562 1000 1778
ρ(Ωm)

**Figure 1.**

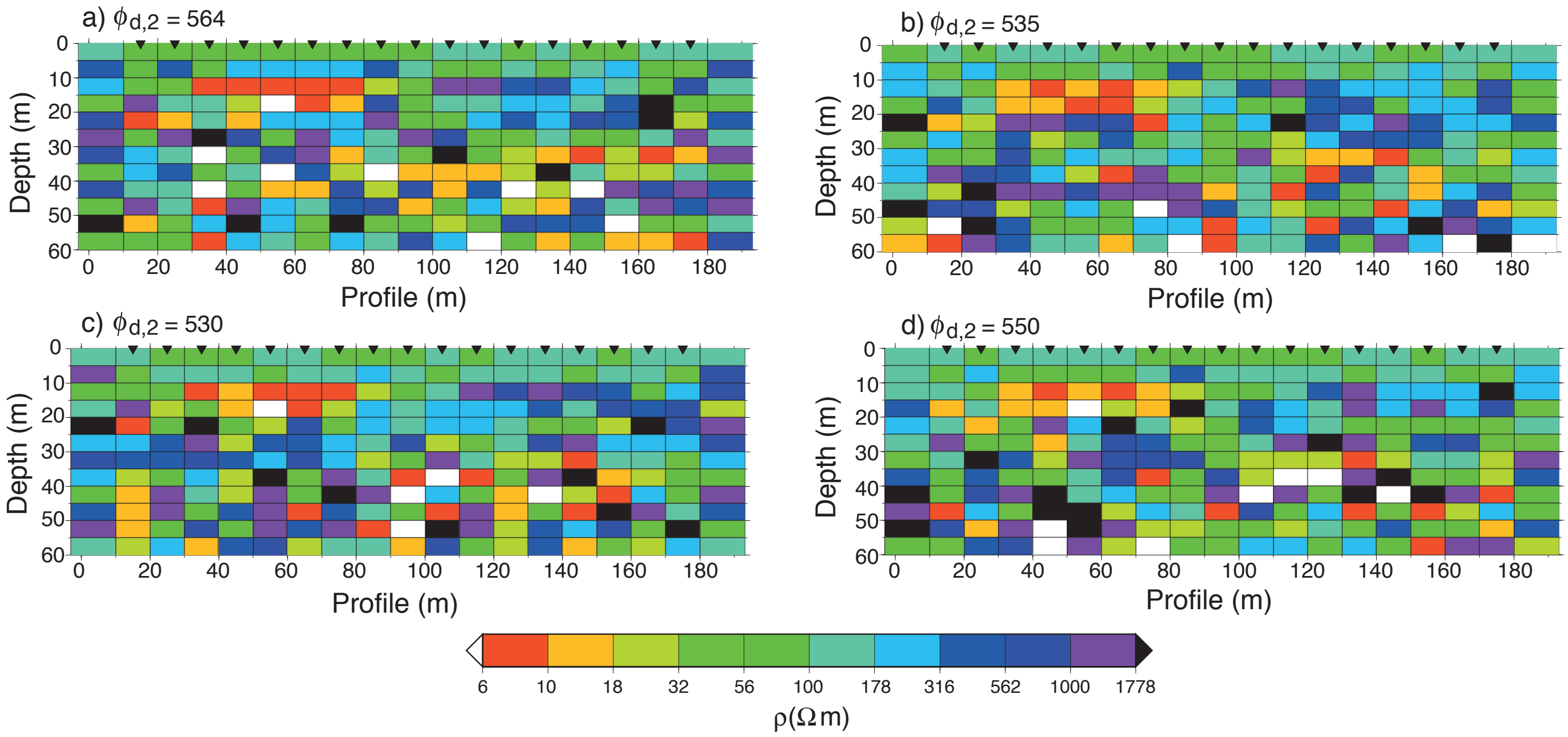
a) $\phi_{d,2}$ = 564
b) $\phi_{d,2}$ = 535
c) $\phi_{d,2}$ = 530
d) $\phi_{d,2}$ = 550
Depth (m)
Profile (m)
0 10 20 30 40 50 60
0 20 40 60 80 100 120 140 160 180
6 10 18 32 56 100 178 316 562 1000 1778
ρ(Ωm)

**Figure 2.**

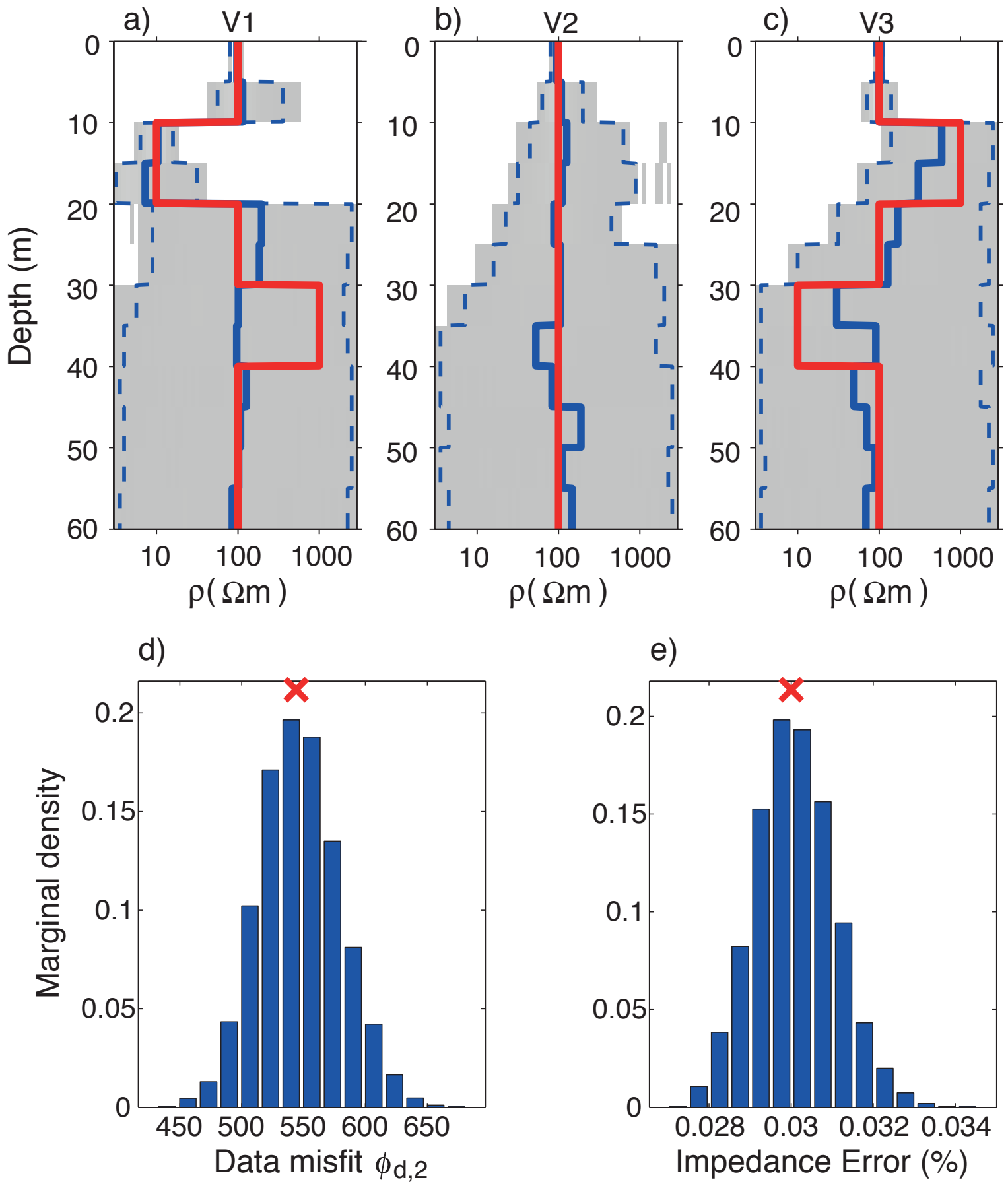

**Figure 3.**

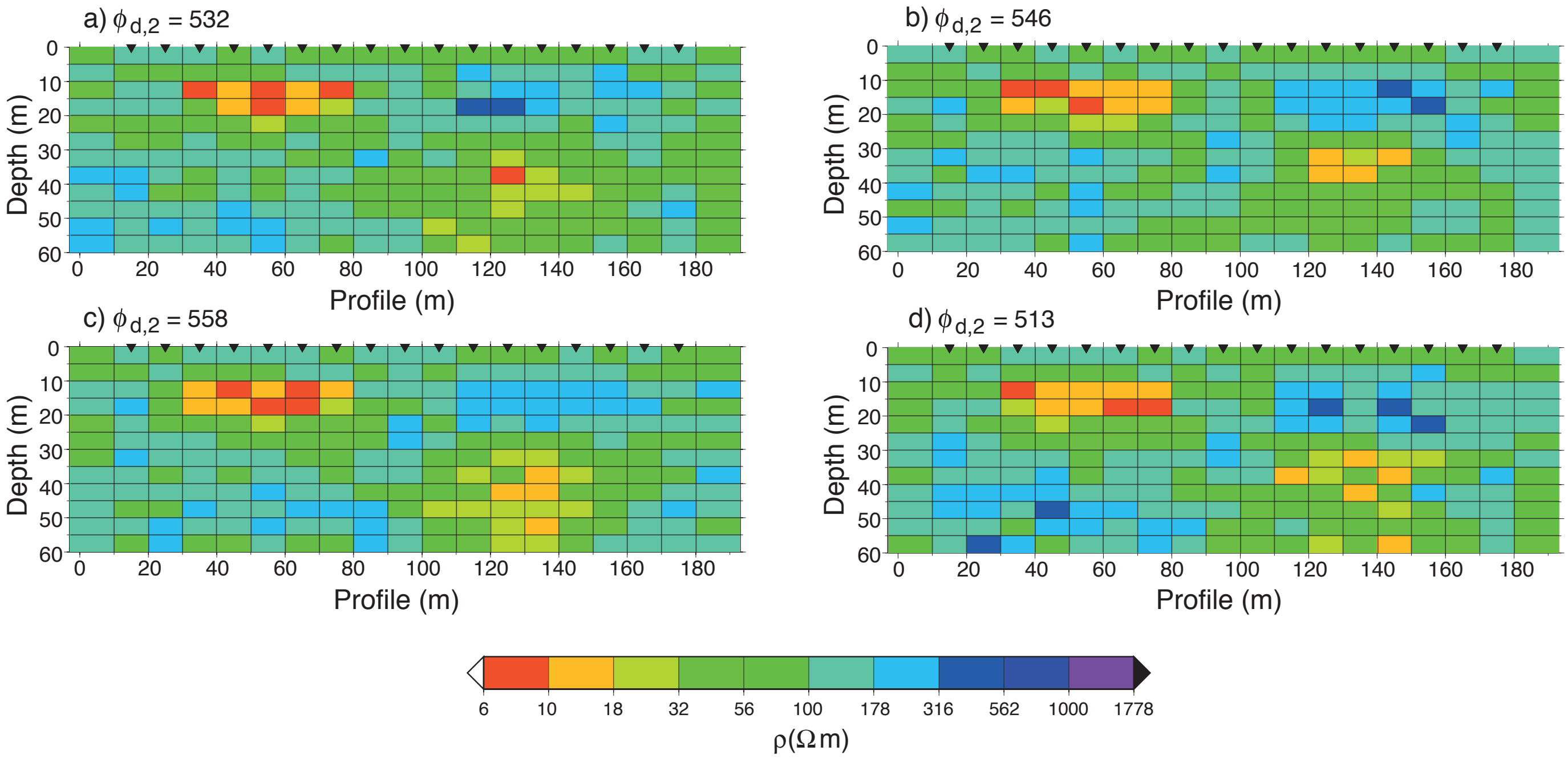

**Figure 4.**

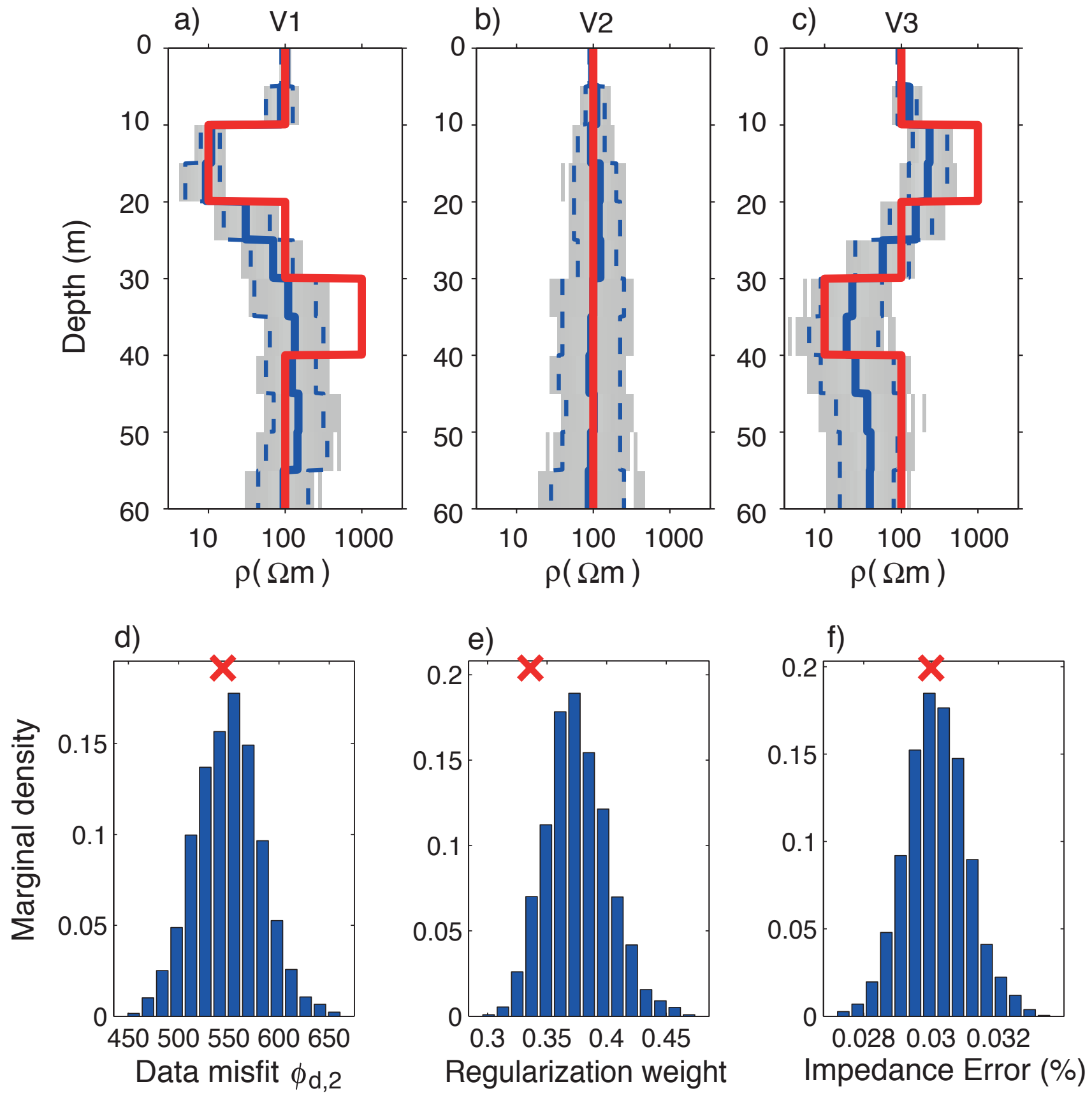

**Figure 5.**

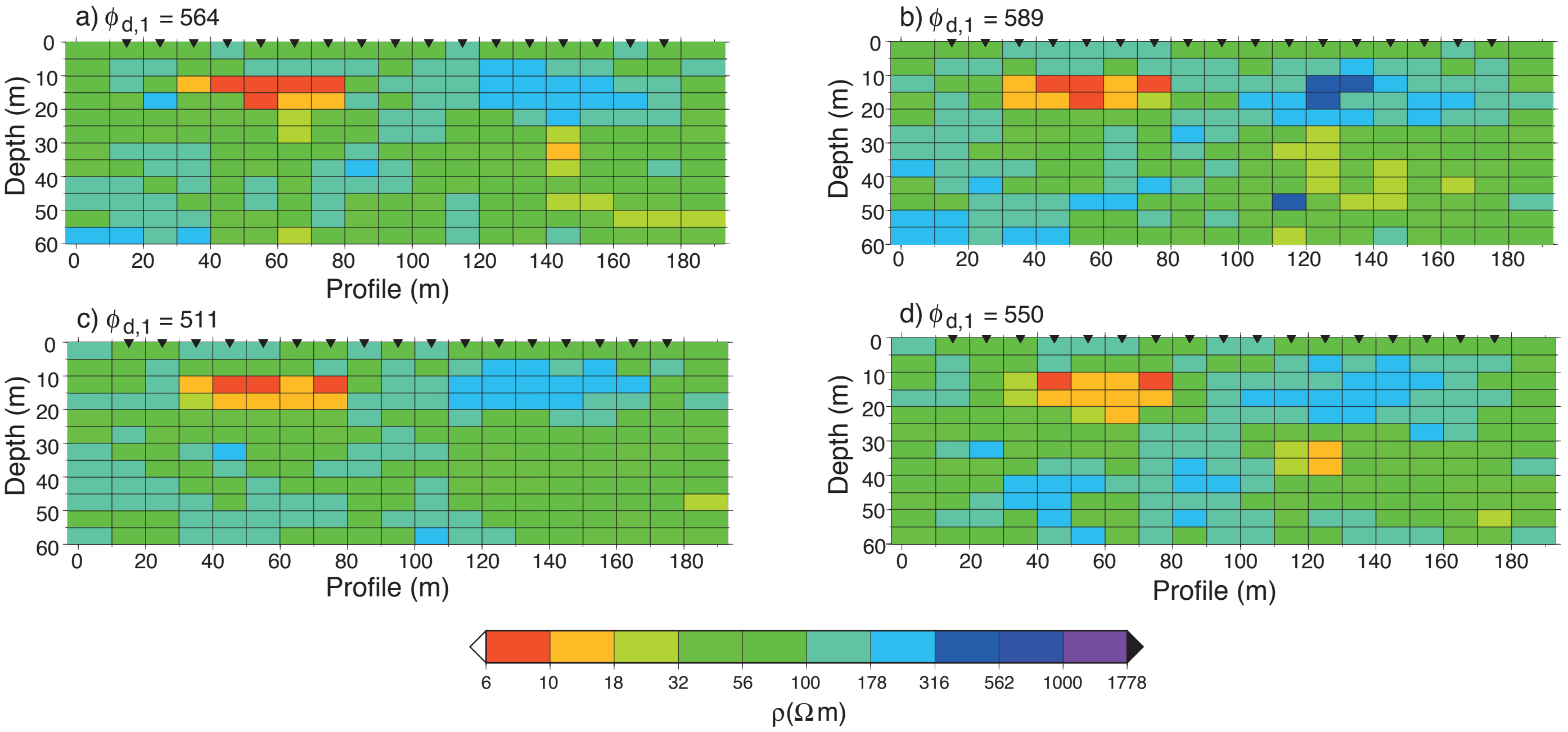

**Figure 6.**

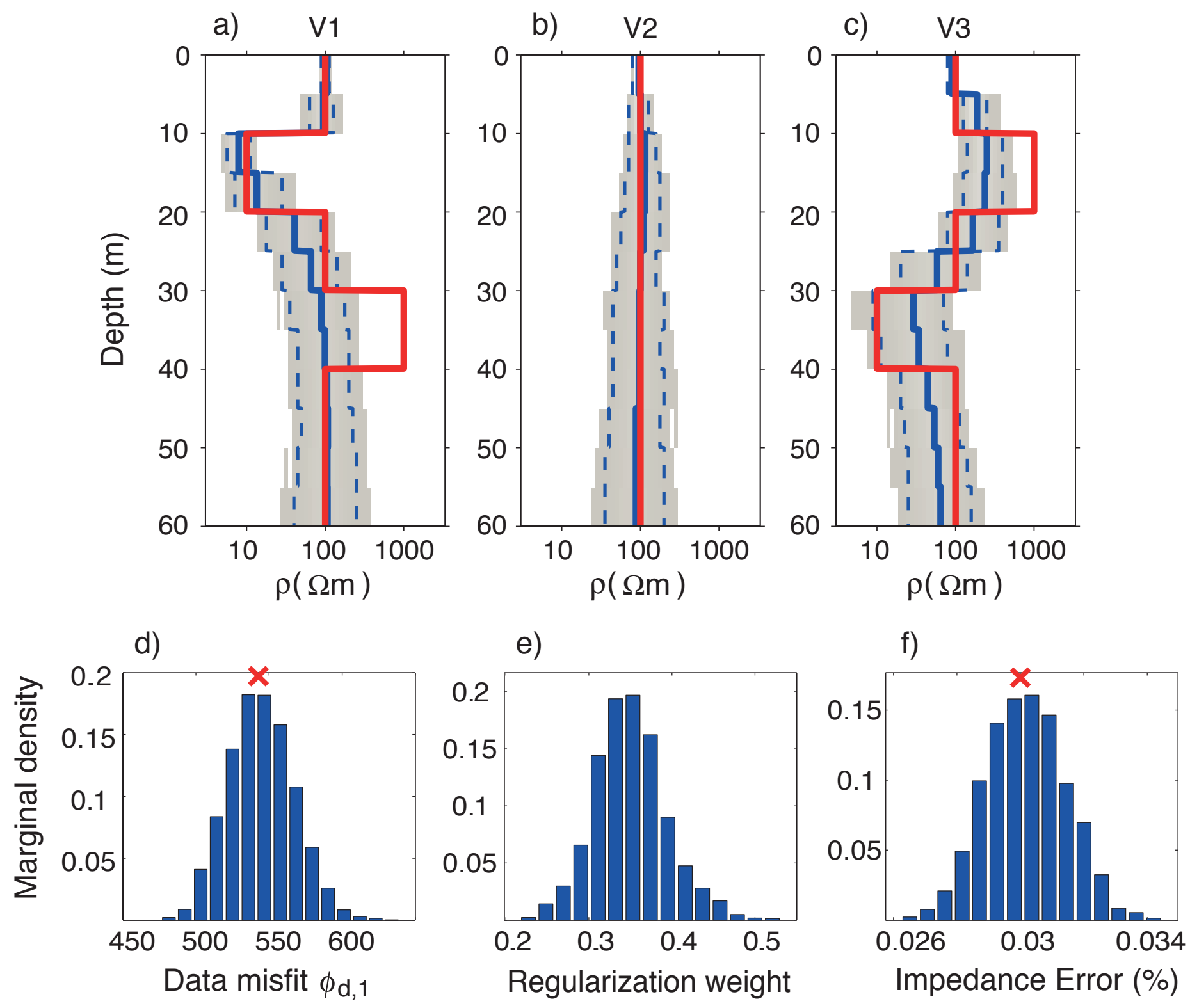

**Figure 7.**

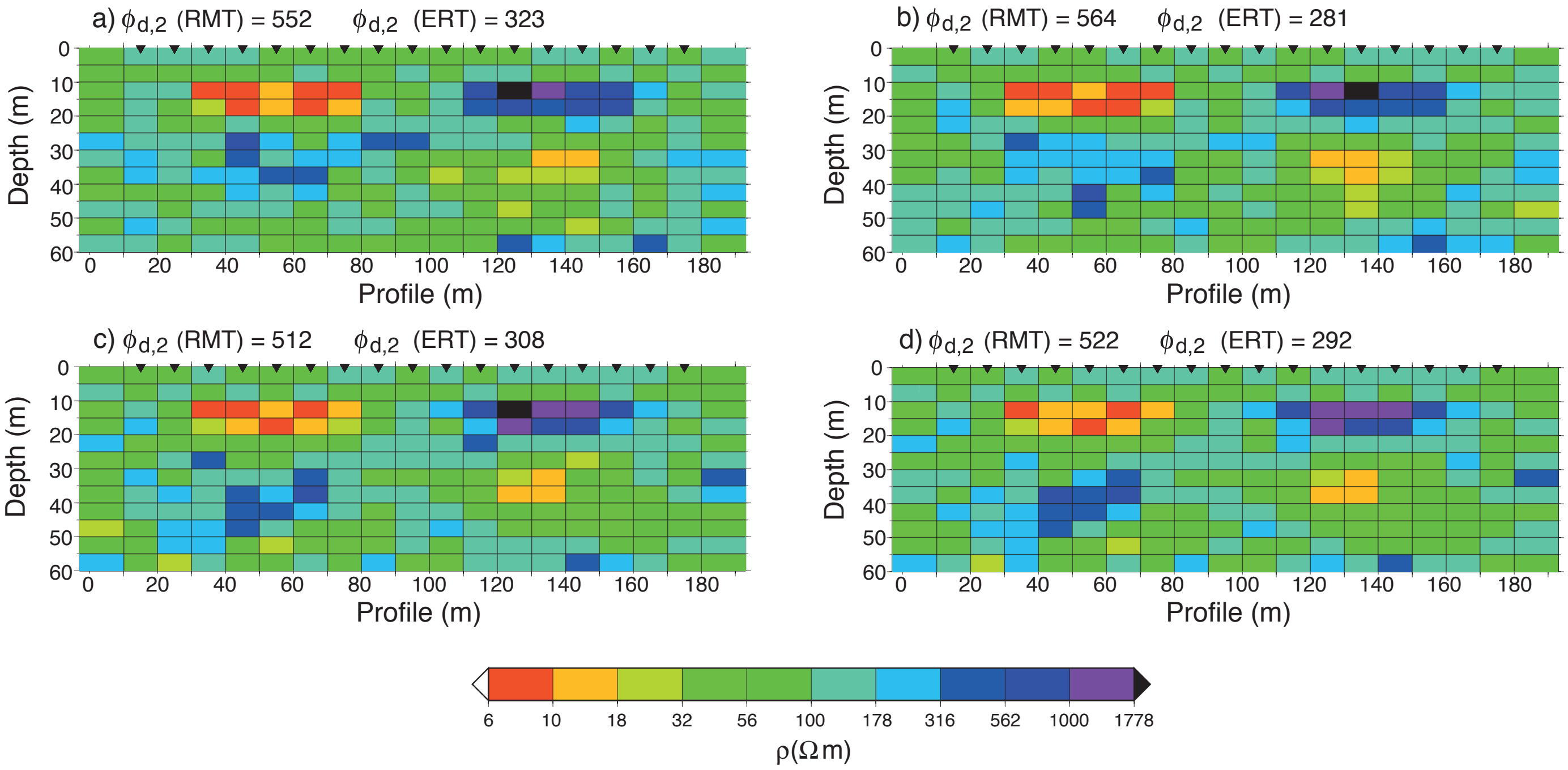

a) $\phi_{d,2}$ (RMT) = 552 $\phi_{d,2}$ (ERT) = 323
b) $\phi_{d,2}$ (RMT) = 564 $\phi_{d,2}$ (ERT) = 281
c) $\phi_{d,2}$ (RMT) = 512 $\phi_{d,2}$ (ERT) = 308
d) $\phi_{d,2}$ (RMT) = 522 $\phi_{d,2}$ (ERT) = 292
Depth (m)
Profile (m)
0 10 20 30 40 50 60
0 20 40 60 80 100 120 140 160 180
6 10 18 32 56 100 178 316 562 1000 1778
ρ(Ω m)

**Figure 8.**

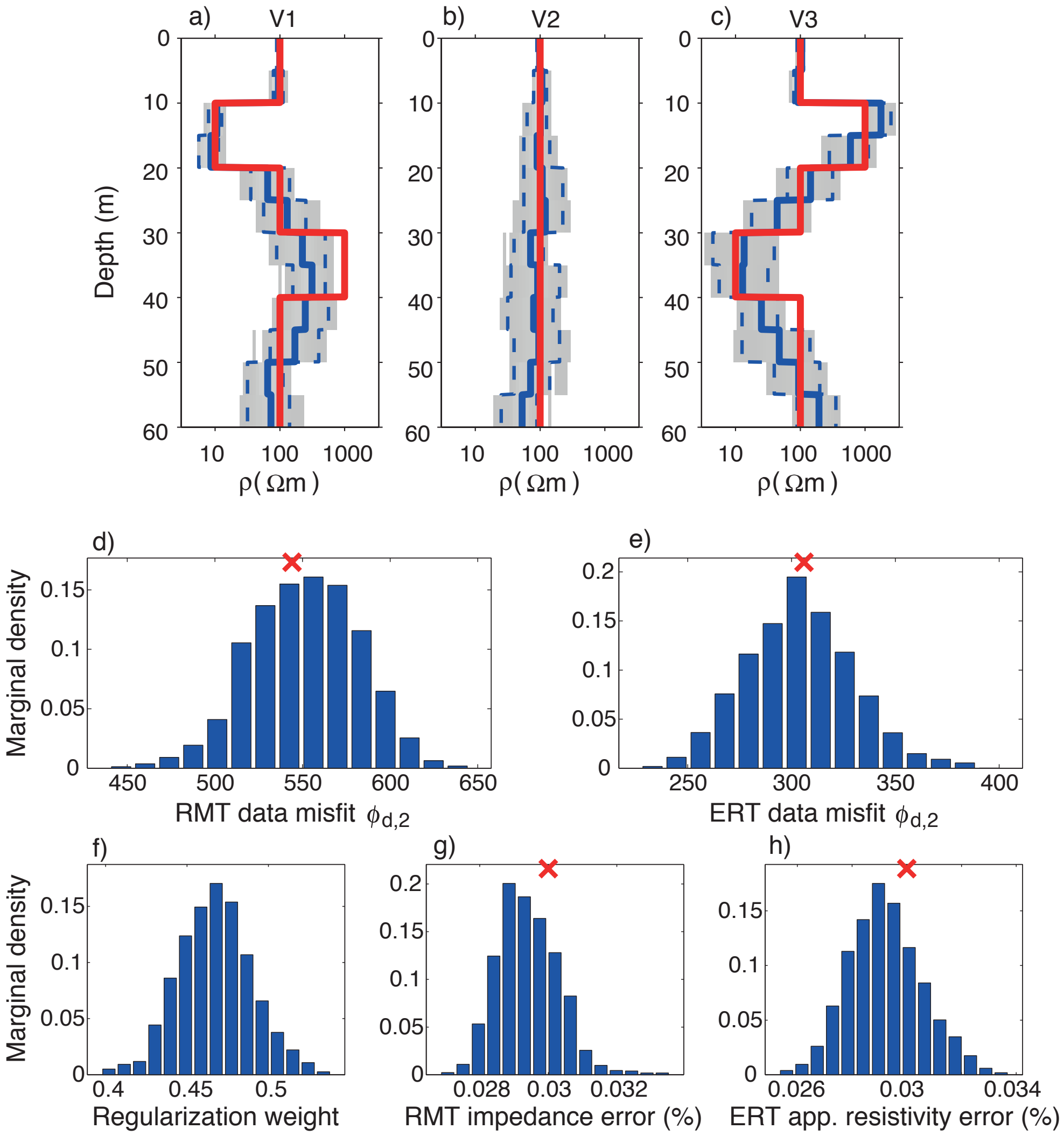

**Figure 9.**

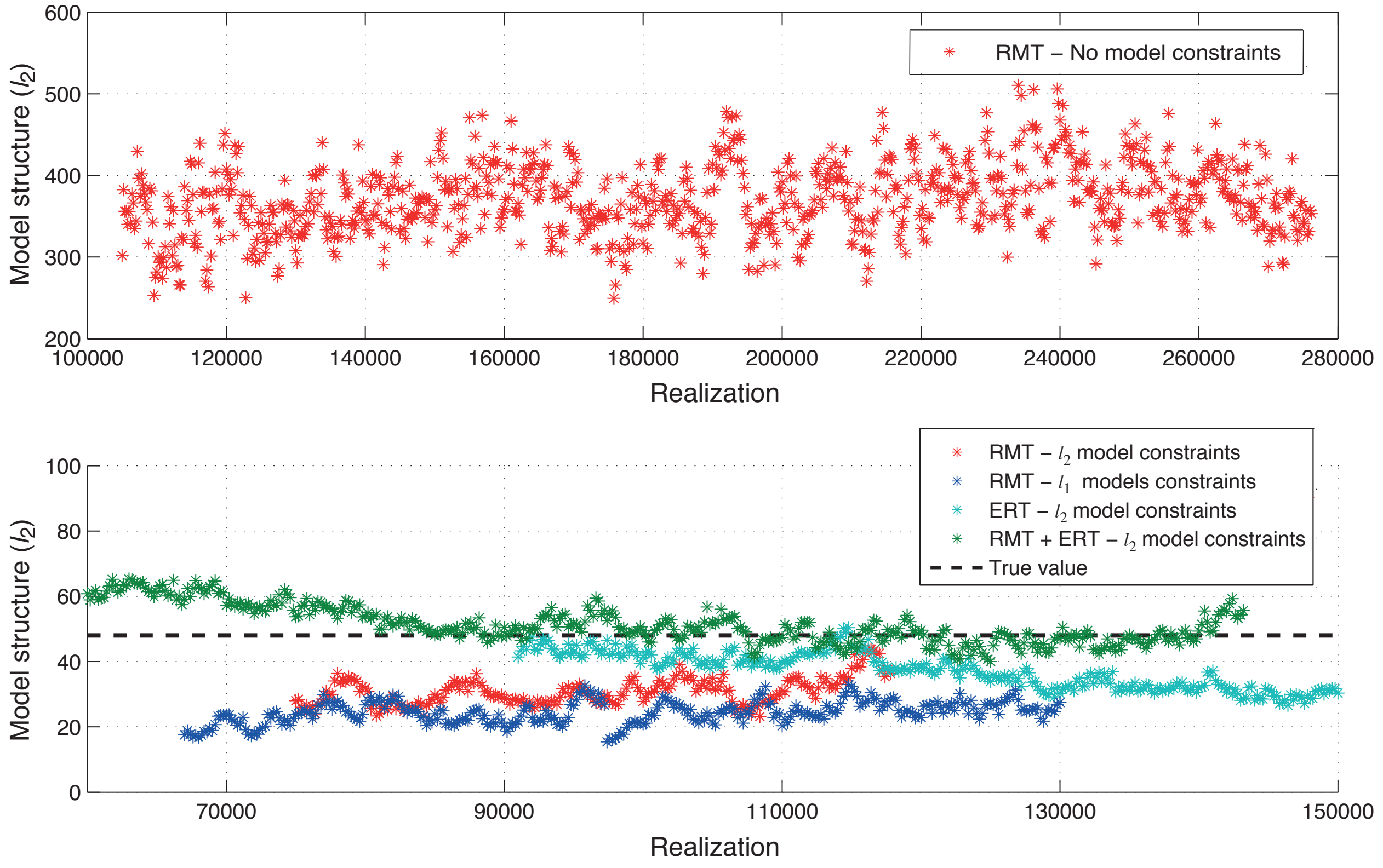

**Figure 10.**

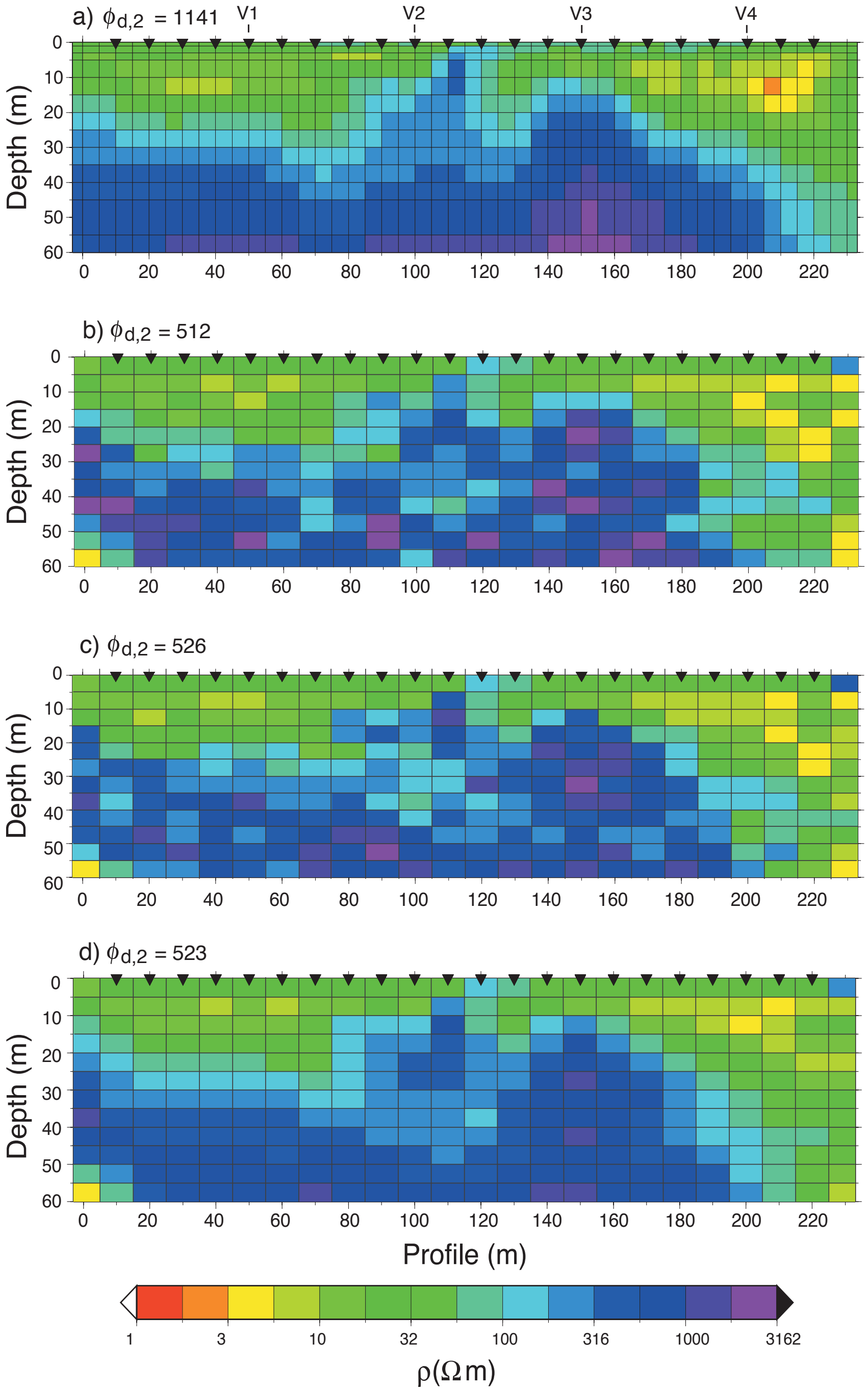

**Figure 11.**

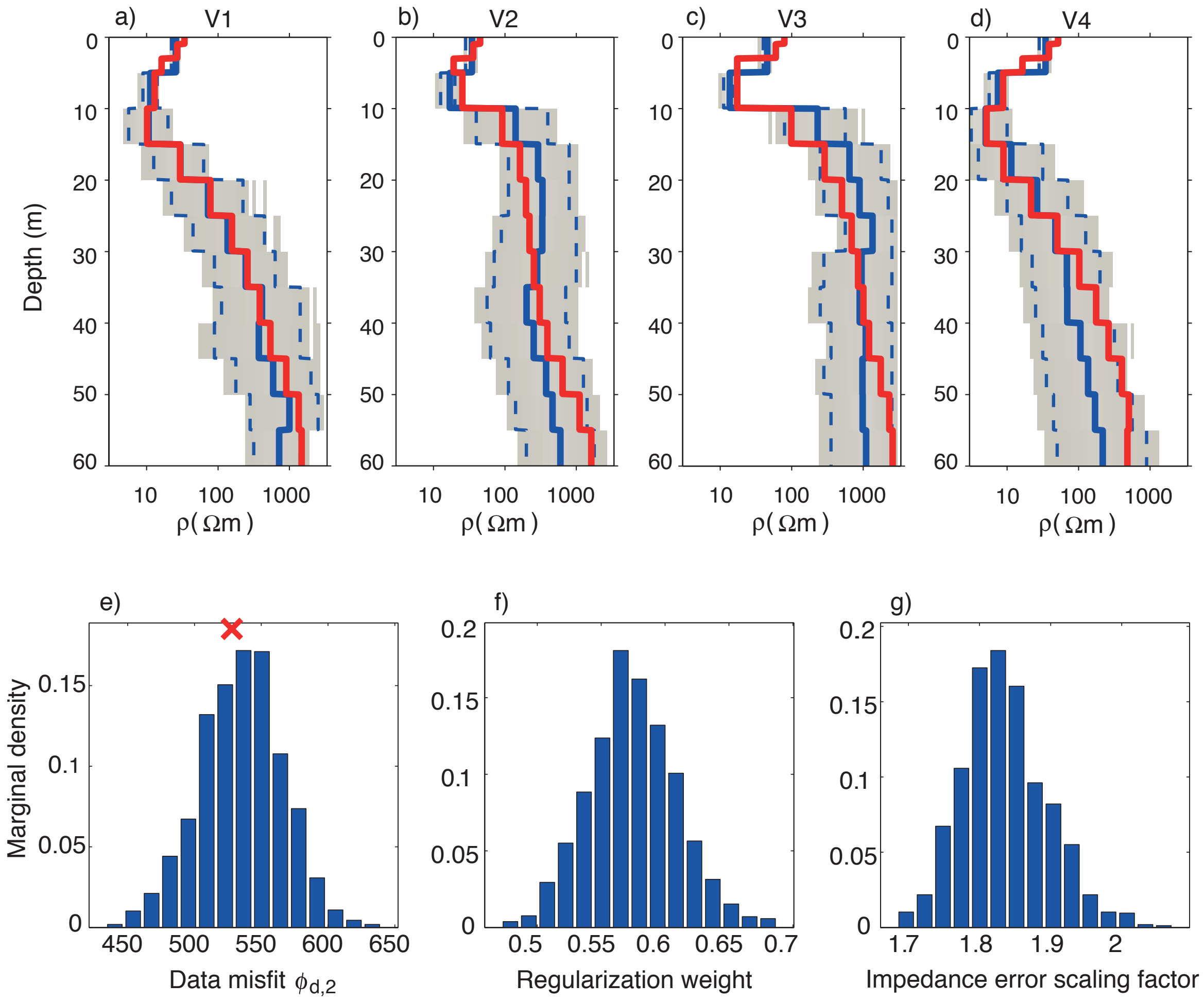

**Figure 12.**